\documentclass[12pt,preprint]{aastex}
\slugcomment{{\sc Accepted to ApJ:} January 10, 2011} 
\usepackage{rotating}
\usepackage{natbib}
\usepackage{lscape}
\usepackage{color}
\tighten

\newcommand{\teff}{\mbox{$T_{eff}$}}

\newcommand{\mjup}{\mbox{$M_{Jup}$}}
\newcommand{\msun}{\mbox{$M_\odot$}}

\newcommand{\mic}{\mbox{$\mu$m}}
\newcommand{\app}{\mbox{$\sim$}}

\newcommand{\pp}{\mbox{$\pm$}}

\newcommand{\dg}{\mbox{$^\circ$}}
\newcommand{\syst}{\mbox{CD$-$35~2722}}
\newcommand{\cand}{\mbox{CD$-$35~2722~B}}
\newcommand{\prim}{\mbox{CD$-$35~2722~A}}

\newcommand{\etal}{et al.}

\newcommand{\eg}{e.g.}

\newcommand{\Lbol}{\mbox{$L_{bol}$}}
\newcommand{\Teff}{\mbox{$T_{\rm eff}$}}
\newcommand{\logg}{\mbox{$\log(g)$}}

\makeatletter
\newcommand{\Rmnum}[1]{\expandafter\@slowromancap\romannumeral #1@}
\makeatother

\begin{document}

\title {The Gemini NICI Planet-Finding Campaign: \\
Discovery of a Substellar L Dwarf Companion to the \\
Nearby Young M Dwarf CD$-$35 2722\footnotemark}

\author {Zahed Wahhaj\altaffilmark{1}, 
Michael C. Liu\altaffilmark{1},  
Beth A. Biller\altaffilmark{1},
Fraser Clarke\altaffilmark{2}, 
Eric L. Nielsen\altaffilmark{3},
Laird M. Close\altaffilmark{3},
Thomas L. Hayward\altaffilmark{4},
Eric E. Mamajek\altaffilmark{5},
Michael Cushing\altaffilmark{6},
Trent Dupuy\altaffilmark{1},
Matthias Tecza\altaffilmark{2},
Niranjan Thatte\altaffilmark{2},
Mark Chun\altaffilmark{1},
Christ Ftaclas\altaffilmark{1},
Markus Hartung\altaffilmark{4},
I.\ Neill Reid\altaffilmark{7},
Evgenya L. Shkolnik\altaffilmark{8},
Silvia H. P. Alencar\altaffilmark{9},
Pawel Artymowicz\altaffilmark{10},
Alan Boss\altaffilmark{8},
Elisabethe de Gouveia Dal Pino\altaffilmark{11},
Jane Gregorio-Hetem\altaffilmark{11},
Shigeru Ida\altaffilmark{12},
Marc Kuchner\altaffilmark{13},
Douglas N. C. Lin\altaffilmark{14}
and Douglas W. Toomey\altaffilmark{15}
}

\altaffiltext{1} {Institute for Astronomy, University of Hawaii, 2680 Woodlawn Drive, Honolulu, HI 96822}
\altaffiltext{2} {Department of Astronomy, University of Oxford, DWB, Keble Road, Oxford OX1 3RH, U.K. }
\altaffiltext{3} {Steward Observatory, University of Arizona, 933 North Cherry Avenue, Tucson, AZ 85721}
\altaffiltext{4} {Gemini Observatory, Southern Operations Center, c/o AURA, Casilla 603, La Serena, Chile}
\altaffiltext{5} {University of Rochester, Department of Physics \& Astronomy, Rochester, NY 14627-0171, USA} 
\altaffiltext{6} {Jet Propulsion Laboratory, MS 169-506, 4800 Oak Grove Drive, Pasadena, CA 91109}
\altaffiltext{7} {Space Telescope Science Institute, 3700 San Martin Drive, Baltimore, MD 21218}
\altaffiltext{8} {Department of Terrestrial Magnetism, Carnegie Institution of Washington, 5241 Broad Branch Road, NW, Washington, DC 20015}
\altaffiltext{9} {Departamento de Fisica - ICEx - Universidade Federal de Minas Gerais, Av. Antonio Carlos, 6627, 30270-901, Belo Horizonte, MG, Brazil}
\altaffiltext{10} {University of Toronto at Scarborough, 1265 Military Trail, Toronto, Ontario M1C 1A4, Canada}
\altaffiltext{11} {Universidade de Sao Paulo, IAG/USP, Departamento de Astronomia, Rua do Matao, 1226, 05508-900, Sao Paulo, SP, Brazil}
\altaffiltext{12} {Tokyo Institute of Technology, 2-12-1 Ookayama, Meguro-ku, Tokyo 152-8550}
\altaffiltext{13} {NASA Goddard Space Flight Center, Exoplanets and Stellar Astrophysics Laboratory, Greenbelt, MD 20771}
\altaffiltext{14} {Department of Astronomy and Astrophysics, University of California, Santa Cruz, CA, USA}
\altaffiltext{15} {Mauna Kea Infrared, LLC, 21 Pookela St., Hilo, HI 96720}

 \begin{abstract}
 We present the discovery of a wide (67~AU) substellar companion to the
 nearby (21~pc) young solar-metallicity M1~dwarf CD$-$35~2722, a member
 of the $\approx$100~Myr AB Doradus association.
 Two epochs of astrometry from the NICI Planet-Finding Campaign confirm
 that \cand\ is physically associated with the primary star.
 Near-IR spectra indicate a spectral type of L4\pp1 with a moderately
 low surface gravity, making it one of the coolest young companions
 found to date.
 The absorption lines and near-IR continuum shape of \cand\ agree
 especially well the dusty field L4.5~dwarf 2MASS~J22244381$-$0158521,
 while the near-IR colors and absolute magnitudes match those of the
 5-Myr old L4 planetary-mass companion, 1RXS~J160929.1$-$210524~b.
 Overall, \cand\ appears to be an intermediate-age benchmark for
 L-dwarfs, with a less peaked $H$-band continuum than the youngest
 objects and near-IR absorption lines comparable to field objects.
 We fit Ames-Dusty model atmospheres to the near-IR spectra and find
 $T_{eff} = $1700--1900~K and $\log(g) = $4.5$\pm$0.5.
 The spectra also show that the radial velocities of components~A
 and~B agree to within \pp10~km/s, further confirming their physical
 association.
 Using the age and bolometric luminosity of \cand, we derive a mass of
 31\pp8~\mjup\ from the Lyon/Dusty evolutionary models.
 Altogether, young late-M to mid-L type companions appear to be
 overluminous for their near-IR spectral type compared to field
 objects, in contrast to the underluminosity of young late-L and
 early-T dwarfs.

\end{abstract}

\footnotetext[1]{Based on observations obtained at the Gemini
  Observatory, which is operated by the Association of Universities for
 Research in Astronomy, Inc., under a cooperative agreement with the
  NSF on behalf of the Gemini partnership: the National Science
  Foundation (United States), the Science and Technology Facilities
  Council (United Kingdom), the National Research Council (Canada),
  CONICYT (Chile), the Australian Research Council (Australia),
  Minist\'{e}rio da Ci\^{e}ncia e Tecnologia (Brazil) and Ministerio de
  Ciencia, Tecnolog\'{i}a e Innovaci\'{o}n Productiva (Argentina).}

\section {Introduction}

Currently, more than 440 extrasolar planets have been discovered by
radial velocity and transit techniques, orbiting at relatively small
separations ($< $ 5~AU) from their parent stars
\citep[e.g.][]{2006ApJ...646..505B, 2008PhST..130a4001M,
  2009A&A...493..639M}. In contrast, direct-imaging surveys are
sensitive to Jupiter-mass planets at separations of $\gtrsim$ 10 AU
around nearby stars (\app\ 25 pc) when they are young
($\lesssim$~100~Myrs) \citep[e.g.][]{2007ApJS..173..143B,
  2007ApJ...670.1367L, 2010ApJ...717..878N}. So far, these searches have
mostly yielded brown dwarf or planetary-mass objects at large
separations \citep{2005A&A...438L..29C, 2006ApJ...649..894L,
  2008ApJ...689L.153L, 2009ApJ...707L.123T}. Spectroscopic studies of
the atmospheres of the brighter companions are revealing their
temperatures, gravities and masses \citep{2009ApJ...704.1098L,
  2010arXiv1001.0470B}. However, even more exciting discoveries of
planetary-mass objects at 5-120 AU separations are beginning to come
forth \citep{2008Sci...322.1348M, 2008Sci...322.1345K,
  2010arXiv1006.3314L}. 
Since these companions have separations that allow us to isolate their
photons from those of the primary, we can begin to characterize the
atmospheres of the companions discovered by direct imaging \citep{2010ApJ...723..850B}.

Since December 2008, we have been conducting the Gemini NICI
Planet-Finding Campaign to find and characterize planets through direct
imaging (Liu \etal\ 2010). The Campaign is a three-year program at the
Gemini-South 8.1-m Telescope to directly image and characterize
gas-giant planets around 300 nearby stars. The NICI instrument was
specifically designed for this purpose
\citep{2003IAUS..211..521F,2008spie.7015E..49C}. It is equipped with an
85-element curvature adaptive optics system, a Lyot coronagraph, and a
dual-camera system capable of simultaneous spectral difference imaging
\citep[SDI]{1999PASP..111..587R} on and off the 1.6~\mic\ methane
feature seen in cool ($<$1400 K) substellar objects
\citep[e.g.][]{1997ApJ...491..856B, 2003IAUS..211...41B}.
The two cameras are equipped with Aladdin II InSb detectors and have
18.43$''$ x 18.43$''$ field of view and a plate scale of 18~mas/pixel (milli-arcseconds/pixel).
NICI can also be used in a fixed Cassegrain Rotator mode for Angular
Differential Imaging
\citep[ADI;][]{2004Sci...305.1442L,2005JRASC..99..130M}, where the
telescope image rotator is turned off, so that the sky rotates on the
detector while the telescope and instrument PSF (point spread function)
artifacts remain fixed (See~\S~2.1.1).

In an earlier paper we presented the discovery of PZ Tel B, one of few
young substellar companions directly imaged at orbital separations
similar to those of giant planets in our own solar system
\citep{2010ApJ...720L..82B}. Here we present the discovery of \cand , 
one of the coolest young ($\sim$~100~Myr) close-separation ($<$~100~AU) brown dwarf 
companions to a star directly-imaged to date. \citet{2008hsf2.book..757T}
identify \prim\ as a member of the AB Doradus association, based on its
common space motion and consistent galactic location with the other
association members. They derive a kinematic distance estimate of \app\
24~pc (See~\S~3.1). The eponymous star of the AB Doradus association itself appears to be coeval
with the Pleiades, which is thought be younger than 150 Myrs \citep{2004ApJ...614..386B, 2004apj...604..272B, 2007MNRAS.377..441O},  
and other studies conclude that AB Dor is approximately 50 Myr or older \citep{2004ApJ...613L..65Z,2007ApJ...665..736C}. 
Thus, we adopt an age of 100\pp50~Myr for \prim\ based on its membership in the AB Dor group.

\section {Observations and Data Reduction}

\subsection{Gemini South/NICI imaging}

\subsubsection{Campaign Observing Modes and Data Reduction \label{sec:reduc}}

We use two observing sequences for each target in the NICI Campaign to
optimize our sensitivity to both methane-bearing and non-methane bearing
companions. Both sequences employ the ADI technique
  \citep{2004Sci...305.1442L,2005JRASC..99..130M}.
Exposure times per frame are kept around 1 minute so that the noise is
background-limited, and the total time spent on detector readouts is
small. Observing hour angles are chosen such that the rotation rate of
the sky is less than 1 degree per minute to avoid excessive smearing
during an individual exposure. The total sky rotation over the observing
sequence is also required to be greater than 15 degrees, to limit flux
loss by self-subtraction of companions at small separations
\citep[e.g.][]{2009AIPC.1094..425B}.

To search for close-in methane-bearing planets, we combine the Angular
and Spectral Difference Imaging methods into a single unified sequence
(``ASDI''). In this observing mode, a 50-50 beam splitter in NICI
divides the incoming light between the off ($\lambda$ = 1.578~\mic) and
on-methane ($\lambda$ = 1.652~\mic) moderate-bandwidth ($\Delta
\lambda$/$\lambda$=4\%) filters which pass the light into the two
imaging cameras, henceforth designated the ``blue'' and ``red'' channels
respectively. The two cameras are read out simultaneously for each
exposure. The red channel image is subtracted from the blue channel
image to remove the nearly identical quasi-static speckles in the two
channels. The resulting difference image has drastically improved
sensitivity to methane-bearing objects, which are expected to be much
fainter in the red channel (methane-absorbed) than in the blue channel.
In the ASDI mode, we take 45 1-minute
frames.

At larger separations ($\gtrsim$~1.5\arcsec),
our sensitivity is limited by throughput, not by speckle noise. Thus we
also obtain an ADI sequence without the 50-50 beam splitter, sending all
the light to the blue channel, in order to achieve maximum sensitivity.
In this mode, we take 20 1-minute frames using the standard $H$-band
filter, which is roughly four times wider than the 4\% methane on/off
filters.

The initial image processing steps in the NICI Campaign data reduction
pipeline were applied to all raw images. After bad pixel removal and
flat-fielding, distortion corrections were applied. To determine the
distortion, we acquired images of a rectangular pinhole grid placed
upstream of the NICI AO system.
Deviations of the imaged pinhole (spot) centers from a rectangular grid
were used to measure the optical distortion of the blue and red
channels. First, we fit a model rectangular grid of points to the spot
centroids, allowing for rotation, change in grid spacing (different in x
and y directions) and translation. Next, we found the 2nd order
transformation in the coordinates of the observed spot centers that
match the coordinates of the fitted model grid.
The RMS in the difference between the distortion-corrected spot
centroids and the rectangular model grid centroids was 0.1 pixels.
Distortion-corrected images were then constructed by replacing the
values at the original coordinates with interpolated values at the
transformed coordinates using cubic interpolation.

In NICI data, the translucent focal plane mask causes arc-like features
to appear near the half transmission radius (r = 0.32$''$) of the
Gaussian-profiled mask, which are dissimilar between the blue and red
channels. Thus, for the ASDI and ADI datasets, we applied a filter to
the images to remove arc-like PSF features centered on the primary star.
For every pixel in the image, we computed the interpolated value at 100
points evenly spaced along an arc centered on the primary star and
extending 30 pixels in azimuth around the chosen pixel. The median of
these 100 values was then subtracted from the pixel value. This
filtering process improves the quality of channel differencing during
the ASDI and ADI reductions.

An additional filter was applied to remove the stellar halo. Each
individual image was smoothed with a Gaussian of 8 pixel FWHM and
subtracted from the original (unsharp masking). Next, all the (unsharp
masked) images were smoothed with a Gaussian of 2 pixel FWHM. With this
2-step smoothing process, all spatial features that are significantly
different in size from the FWHM \app\ 3~pixel NICI PSF were removed from
the images.

\subsubsection{ADI and ASDI Observations}

\begin{figure}[ht] 
  \centerline{
    \includegraphics[height=12cm]{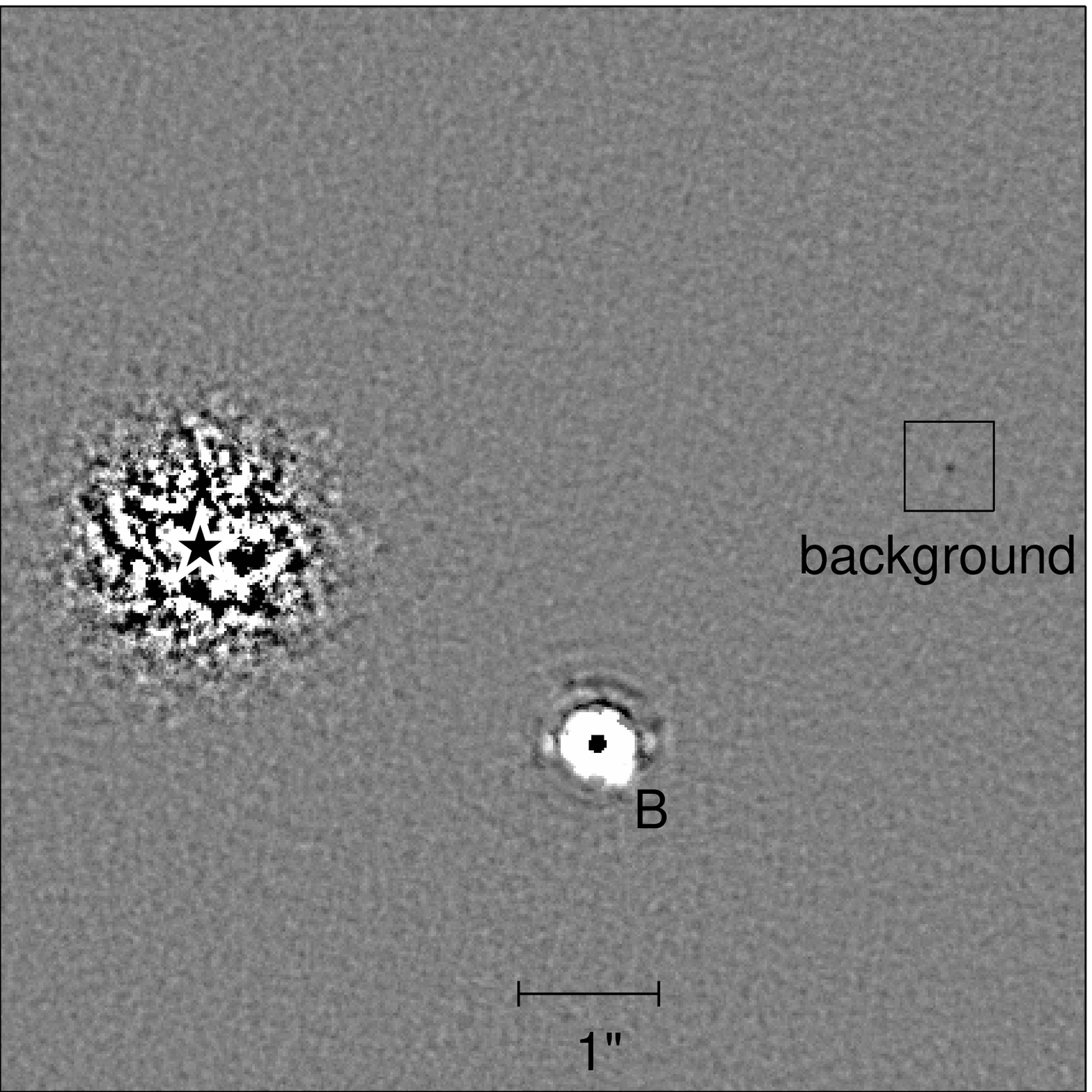}     
  }
  \caption{A cutout from the reduced NICI image of CD-35 2722,
    representing a 45-minute ADI observing sequence (North is up, East
    is left). This image is the sum of the reduced images from the blue
    and red channels (CH$_{4}$ 4\% short and long filters). The stellar position is indicated by a 'star'
    symbol. \cand\ was detected 3.137$''$ away from the primary at a PA
    of 243.10\dg\ on UT Jan 10, 2010. The background object, shown in
    the square, had 5.316$''$ separation and PA=276.32\dg . The
    pixel intensities are shown in linear stretch from $-$4$\sigma$ to
    4$\sigma$, where $\sigma$ is the pixel-to-pixel noise in background.
  }
  \label{cd35}
\end{figure}

\syst\ was first observed on UT January 16, 2009 in the standard
campaign observing modes (ASDI and ADI). The NICI images revealed a
bright companion candidate with separation $\approx$~3$''$. For
astrometric follow-up, we observed the system again on UT January 10,
2010 in the ASDI mode.
Since the companion has separation $>$ 1$''$ and does not have methane,
we reduced the blue and red channel data sets separately using a simple
ADI data reduction method and then added the reduced images together.
For each channel, all the images were registered using the primary star,
which was detected at very high signal to noise through the partially
transmissive focal plane mask. The images were median combined to build
a PSF image. This image consisted only of quasi-static speckles, since
any real objects in the dataset would appear at different PAs in the
individual images (since the rotator is off) and would be removed by the
median process. Next, the PSF image was matched to and subtracted from
each of the 45 individual frames, fitting for the optimal translation
and intensity normalization to match the speckle patterns between
0.65$''$ and 0.78$''$ from the star. This fitting region lies well
outside the mask edge at 0.51$''$.
The final reduced image was constructed by rotating the individual
subtracted images to place North up, registering them using the
previously computed positions of the primary star, and median combining
the stack. The PAs recorded in the FITS header of each frame give the
orientation at the beginning of the exposures, but we calculated and
used the PAs at mid-exposure to align our images. Finally, the reduced
blue and red channel images were added together.
We show a cutout from the final reduced image for the 2010 data in
Figure~\ref{cd35}.

\subsubsection{$JHK_S$ Rotator-On Photometric Observations}

\begin{figure}[ht] 
  \centerline{
    \includegraphics[width=20cm]{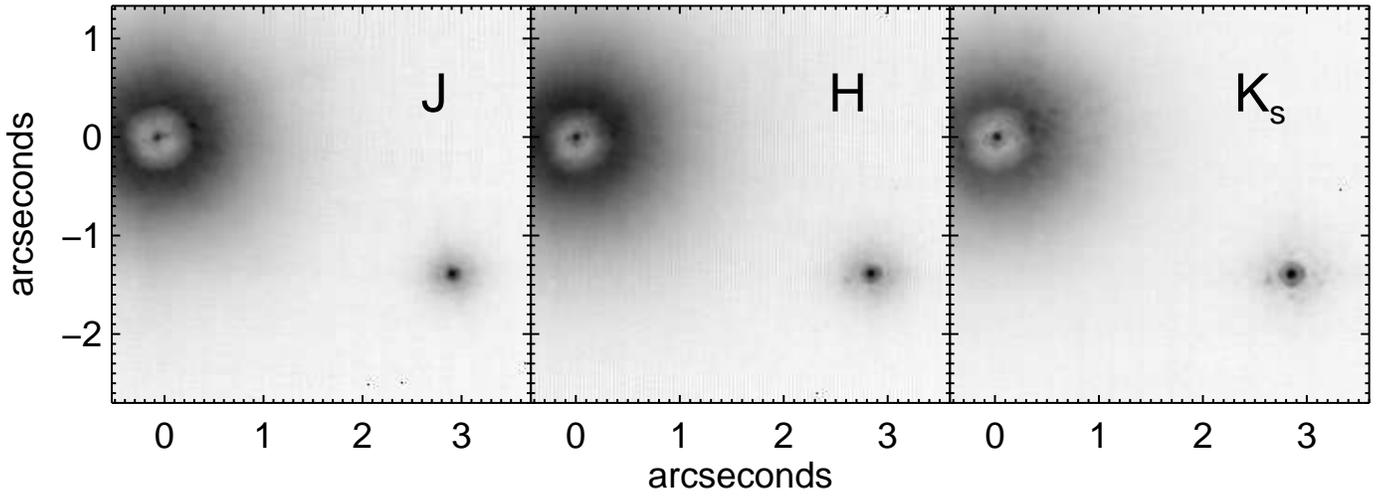}        
  }
  \caption{NICI images of CD-35 2722, obtained on UT March 9, 2009 in
    the $J$, $H$ and $K_S$-band filters. North is up, East is left. The
    primary star, attenuated by the partially transmissive coronagraphic
    mask, is in the upper left area of the panels. \cand\ is in the
    lower right area, roughly 3$''$ away from the primary. }
\label{cd35-jhk}
\end{figure}

Upon detection of the very bright candidate at a separation of 
$\approx$~3$''$, we obtained images in the $JHK_S$ filters on UT March 9, 2009
to measure its near infrared colors. The broad band filters in NICI are
in the MKO (Mauna Kea Observatory) system
\citep{2002PASP..114..169S,2005PASP..117..421T}. The observations were
obtained in the rotator-on mode.
Nine 15-second exposures were taken, imaging the star simultaneously in
both channels, with the $H$ filter in the blue channel and the $K_S$
filter in the red channel. Nine 15-second exposures with the $J$ filter
in the blue channel were taken immediately following these observations.
For the reduction of the $JHK_S$ rotator-on data, we flat-fielded and
registered the frames on the stellar centroid and median combined the
stack of images. Cutouts from the reduced images of \syst\ in the $J$,
$H$ and $K_S$-bands are shown in Figure~\ref{cd35-jhk}.
To determine the $JHK$ band flux ratios of the companion relative to the
primary, we first used simple aperture photometry (3 pixel radius) to
measure the relative fluxes of the companion and the primary star, with
the latter seen under the focal plane mask.

To determine the transmission of NICI's coronagraphic mask, we imaged 
a pair of field stars, 4.2$''$ apart on UT October 31, 2010. The stars, 2MASS J06180157$-$1412573 ($H$=9.46 mag) and 2MASS J06180179$-$1412599  ($H$=11.16 mag)
were chosen so that they could be imaged at the same time without saturating either. Ten images were obtained 
in the $H$ and $K$-bands, with the brighter star under the mask, followed by
10 images with both stars off the mask. In the $J$-band, 4 images were taken in each configuration.  
All images were taken with the Cassegrain rotator on, and thus there was no smearing during
individual exposures. We then computed the ratio of the aperture (3 pixel radius) fluxes of the two stars,
in both the on and off-mask configurations. The off-mask flux ratio divided by the on-mask flux ratio
yields the mask attenuation. We determined that the central attenuation of the 0.32\arcsec\ mask is
6.37\pp0.06, 5.94\pp0.02 and 5.70\pp0.03 magnitudes in the $J$, $H$ and
$K_S$-bands, respectively. The uncertainties are the RMS of the central attenuation measurements 
for the individual images in each filter.

These attenuation factors were applied to the raw flux ratio to arrive
at the final $JHK_S$ flux ratios. We computed the uncertainty in the
flux ratios by including the dispersion in the flux ratios over the data
set and the uncertainties in the focal plane mask transmission.
To convert the $K_S$-band photometry of the companion to the $K$-band
filter (a very small effect), we applied the transformations from 
\citet{2004PASP..116....9S}. Table~\ref{sysprops} lists the final NICI
photometry of \cand. 

\subsection{IRTF/SpeX Imaging}
 
We also obtained images of \syst\ with the infrared guide camera of the facility spectrograph, SpeX, 
of the 3-m NASA Infrared Telescope Facility (IRTF) on UT February 3, 2010 and UT September 24, 2010 in good seeing conditions.
We obtained images in the  $JHK$ bands, using 5 dither positions for each filter.
The IRTF/SpeX images were flat-fielded, sky-subtracted, registered and stacked.
To measure the flux ratio of the primary and the companion we first removed the background 
flux near the companion that comes from the primary. This was achieved by applying the same 
filter used in \S~\ref{sec:reduc} to remove arc-like features in the NICI PSF, except that medians were 
taken over arcs of length 90 pixels, instead of 30. Then flux ratios between the primary and the companion were computed 
using apertures of 6 pixel radii. 
To assess the error in our measurements and any flux loss from the filtering process, 
a hundred simulated companion images were created by scaling down the primary to the flux level
of the secondary and shifting them to the same separation, placed with uniform aziumuthal spacing, 
but at $>$20\dg\ difference in PA ($>$10 pixels away)
compared to the actual companion. Photometry was done on the known positions of the simulated companions
after the filtering process was applied. The RMS in the measurements was taken as the photometric error. 
We also computed the estimated flux loss from the mean of the differences between the known and measured 
photometry of the 100 simulated companions. These loss fractions were $\sim$ 0.03 magnitudes, and we corrected for this loss in each band. 
Even though the two epochs of IRTF measurements agree to better than 0.02 magnitudes, to be conservative we adopt the errors from the 
simulations, and list the mean of the photometry from the two epochs in Table~\ref{sysprops}.

The primary star could be variable and thus use of the non-contemporaneous 
2MASS photometry to derive the magnitudes of \cand\ from our flux ratio
measurements may not be appropriate. Thus we observed a standard star 
in $JHK$ during our 2nd epoch IRTF observations on 
UT September 24, 2010. We measured $JHK$ MKO magnitudes of 
7.85$\pm$0.03, 7.31$\pm$0.05 and 7.07$\pm$0.01 mag for \prim , which agree 
with the 2MASS-derived measurements to within
1$\sigma$ (see Table~\ref{sysprops}). Moreover, other available data do not strongly indicate variability. Optical
monitoring of \prim\ shows variability with an amplitude of $\Delta{V}$
= 0.11--0.12~mag and a period of 1.7~days \citep{2003AcA....53..341P,
  2010arXiv1004.1959M}. This is likely due to rotating spots on the
surface of this young star, in which case the photometric amplitude
should be smaller in the near-IR. 

\subsection {Gemini North/NIFS Spectra \label{sec:nifs}}

Spectroscopic follow-up observations were obtained with NIFS 
\citep{mcgregor03} on the Gemini-North telescope. 
The target was observed with the {\em J, H} and {\em K}
settings of NIFS, which cover 1.15--1.33~$\mu$m,
1.49--1.80~$\mu$m and 1.99--2.40~$\mu$m, respectively, at spectral resolutions of 
$\Delta \lambda$/$\lambda =$~5000--6000. Table~\ref{spectroscopy_observations} gives a log
of the observations. 
The primary star was used as the AO natural guide star, providing
significant correction of the atmospheric turbulence. 
Due to the southern declination of the target (DEC \app\ -36\dg ), the observations were carried out
at high airmasses of 1.8--2.0. 
The average FWHM obtained was 0.15$''$.

\begin{figure}[ht] 
  \centerline{
    \includegraphics[height=5cm]{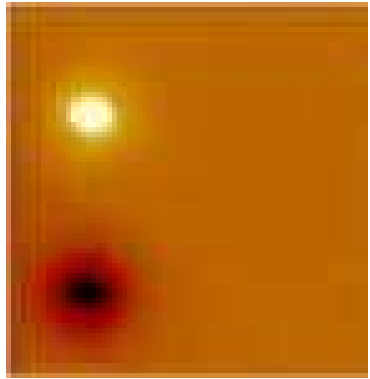}     
  }
  \caption{A NIFS A$-$B image of \cand\ with a 3.2$''\times $3.2$''$
    FOV. North is up, East is left. The primary, not shown in the image,
    lies above and to the left of this region. }
  \label{cd35nifs}
\end{figure}

The data were taken in a 'ABBA'
nodding sequence, with the object dithered by 1.5$''$ between
two positions in the field of view. 
The data were reduced with the Gemini NIFS IRAF
pipeline, following the standard procedures to produce a wavelength
and distortion-corrected data cube. 
The sky background was removed
from A frames by subtracting the negative of the nearest B frame, and
vice versa. Figure~\ref{cd35nifs} shows \cand\ in an A$-$B NIFS image.

Spectra of the object were then
extracted from each cube with the {\tt nfextract} routine, using an
aperture of 0.5$''$. The wavelength solution for the 1-d spectra
was checked against night sky lines (by re-running the reduction
procedure {\em without} the sky subtraction), and re-interpolated 
where necessary. We found small drifts in the wavelength solution
between nights on the order of $\approx$~2~\AA.

Telluric standard stars of spectral type A0V were observed at similar airmass
to the science observations and were reduced in the same way. 
Stellar features (hydrogen lines) were
removed with the IDL package {\tt spextool} \citep{cushing04}. The resulting
telluric spectrum was then used to correct each individual science
spectrum before combining them.

\begin{figure}
  \centerline{
    \hbox{
      \includegraphics[width=18cm]{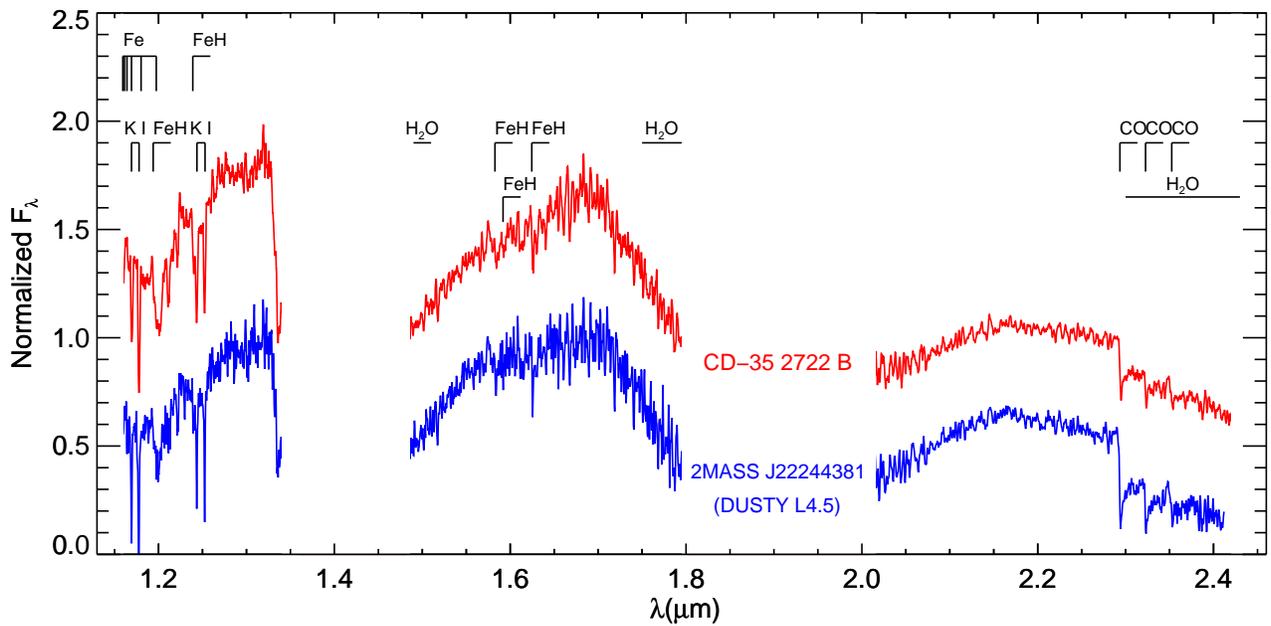}        
    }
  }
  \caption{Normalized $JHK$-band NIFS spectra of \cand (red), plotted along with 
    the dusty L4.5 dwarf 2MASS J22244381$-$0158521 (blue) \citep{2004AJ....127.3553K}. 
    Molecular features and line identifications are from
    \citet{2005ApJ...623.1115C}. }
  \label{nifspec}
\end{figure}

The individual science spectra were scaled and combined using a
3-sigma clipped mean, with the rms per wavelength bin used to estimate 
the noise in the resulting combined spectrum. The combined spectrum
was then multiplied by the filter curves for the Gemini {\em JHK} 
filters, while the total flux in each band was set to the $JHK$ photometry 
measured by NICI. The calibrated $JHK$ spectra are shown in Figure~\ref{nifspec}.

\section {Results}

\subsection {Astrometric Confirmation of \cand }

Two objects were detected in the \syst\ field at separations of
3.172\pp0.005$''$ and 5.312\pp0.011$''$ from \prim\ on UT January 16,
2009 and again at separations of 3.137\pp0.005$''$ and 5.316\pp0.011$''$
on UT January 10, 2010. Since the coronagraphic mask is not completely
opaque, we should be able to measure the position of both the primary
star and the candidate to better than a tenth of a pixel, given the
signal to noise of the detections and the accuracy of our centroiding
algorithm. However, our astrometry is limited by the uncertainty in our
measurement of the image distortion. Comparison of two epochs of dense
star field images shows
the astrometric discrepancy between epochs was \app\ 0.3 pixel at 3$''$ and \app\ 0.6 pixel at 6$''$ 
and we incorporate these errors in our astrometry in Table~\ref{astrom}.

To check whether these astrometric measurements are consistent with
the motions of background objects, we need the proper motion and the distance
to the primary. The proper motions for \prim\ in RA
and DEC are $-$5.6\pp0.9 and $-$56.6\pp0.9 mas/year, respectively \citep{2009yCat.1315....0Z}.

Using the CAPScam/duPont telescope at Las Campanas Observatory \citep{2009PASP..121.1218B}, we obtained 3 epochs of parallax observations separated by 8 months (April 2010, November 2010, December 2010). This yielded a trigonometric parallax of 47\pp3~mas (21.3\pp1.4~pc). Three is the minimum number of observations required to measure a parallax. Thus it is difficult to properly assess the uncertainty in our measurement. So we estimated the error using a Monte Carlo approach, assuming a conservative 1.5~mas precision on each epoch and fitting the parallax and proper motion of 10,000 Monte Carlo realizations of the data. Further observations are in progress and a definitive value will be published in Anglada-Escud\'{e} et al.\ 2010 (in preparation).
 
The mean distance to the AB Doradus group nucleus has been refined
in a new investigation of the moving groups within 100 pc of the Earth \citep[see][]{2010AAS...21545505M}.  
The kinematic analysis therein employs the space motions of only the group members with
Hipparcos parallaxes \citep[][]{2007A&A...474..653V} and the highest quality radial velocity measurements 
to find a distance of 20.6$\pm$2.4~pc for 8 ``nuclear'' members. 
Comparing the AB Dor group space motions and the \prim\ proper motion, we see that the star is very likely also a
"nuclear" AB Doradus member and thus at the same distance. We note that this distance estimate is in agreement 
with our parallax measurement. The predicted radial velocity from the best-fit space motions is 31.5\pp1.4~km/s. 
This is completely consistent with the published measurement of 31.4\pp0.4~km/s \citep{2006A&A...460..695T}.

\begin{figure}[ht] 
  \centerline{
    \vbox{
      \hbox {
        \includegraphics[width=8cm]{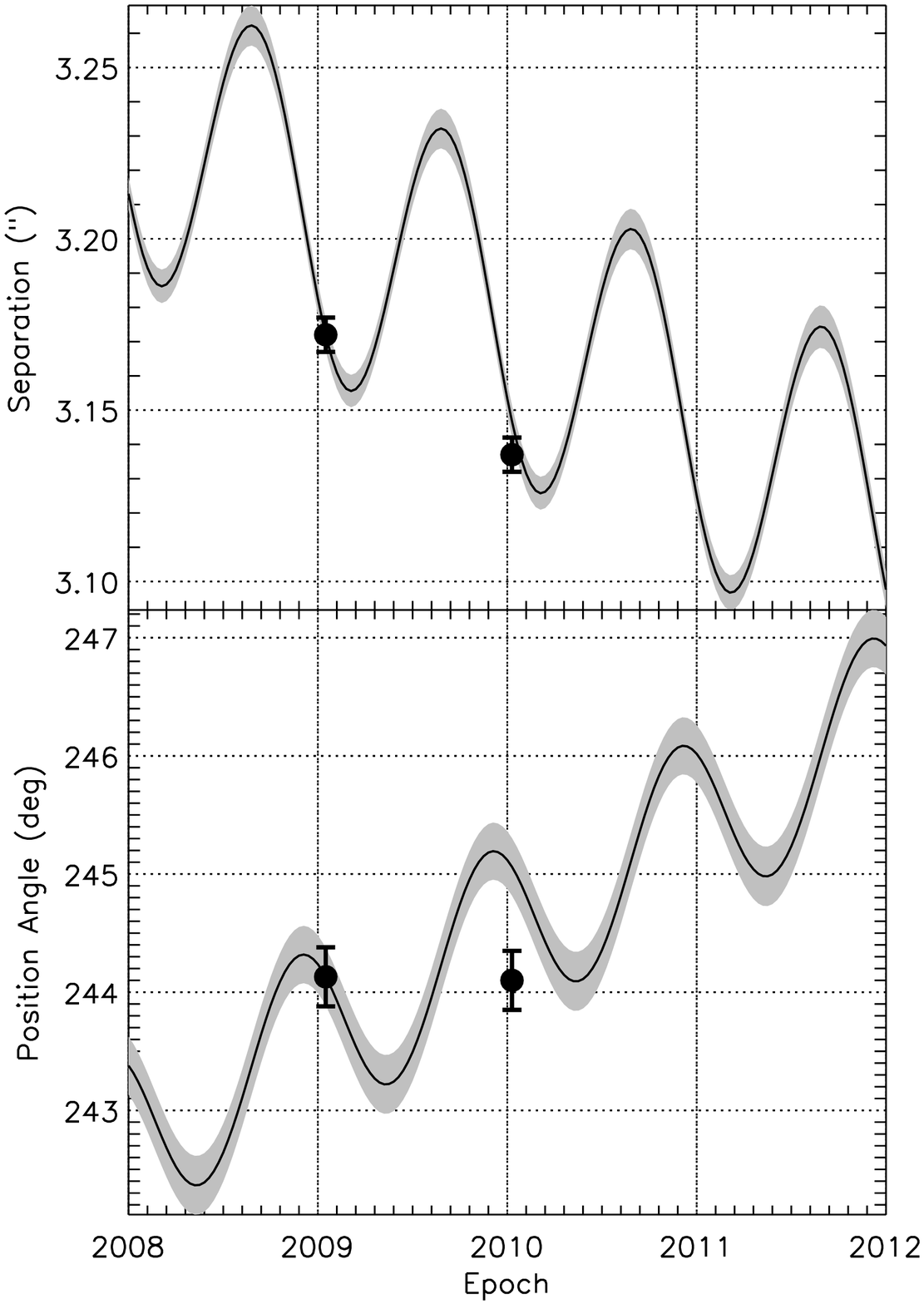}        
        \includegraphics[width=8cm]{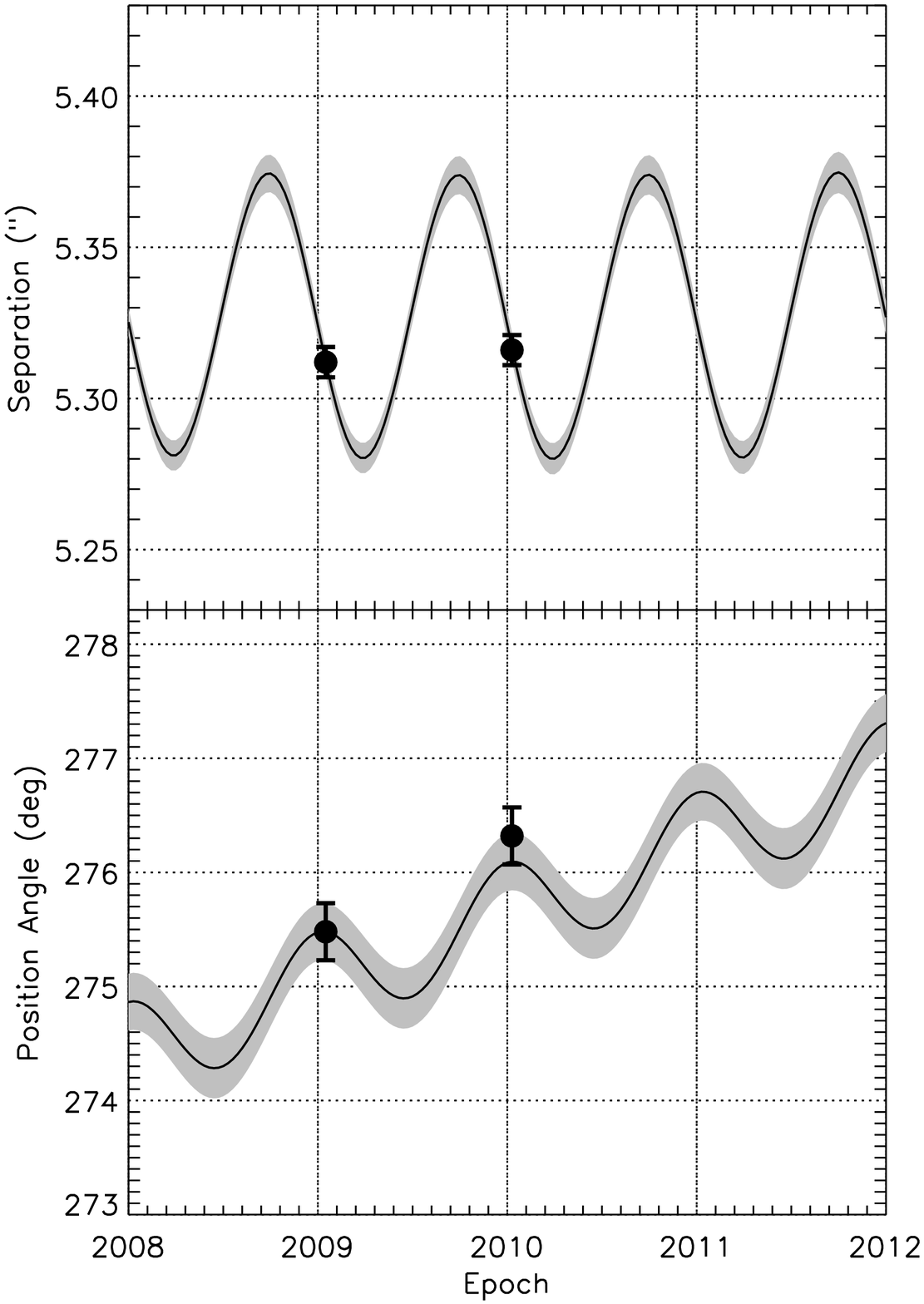}        
      }
    }
  } 
  \caption{Comparisons of the observed motions of the two candidates
    (Left: \cand ; Right: background object) to the expected motion of a
    background object given the first epoch astrometry, the proper
    motion and the distance of \prim . The very small observed change of
    the PA of \cand\ (bottom left panel) differs significantly
    (3$\sigma$) from that expected for a background object. The change
    in the position of the other candidate is consistent with a
    background object.}
  \label{bgndmotion}
\end{figure}

In Figure~\ref{bgndmotion}, we compare the motions of the two objects in
the \syst\ field to the expected motion of a background object given the
first epoch astrometry, the known proper motion and estimated distance
to \prim .
The change in the PA of \cand\ is significantly different (at the
3$\sigma$ level) from that expected of a background object. This
indicates that \cand\ is physically associated with the primary,
although about 1.6~AU/year of orbital motion is required to explain the
relative motion with \prim . In comparison, the second object in the
field at $\approx$~5.3\arcsec\ separation moves as expected for a
background object, thus confirming the joint accuracy of the proper
motion and parallax of \prim\ and our NICI astrometric calibration. (See
also Biller \etal\ 2010.)

Since the viewing angle and eccentricity of companion's orbit are
unknown, we estimate its semi-major axis by applying a statistical
correction from \citet{2010ApJ.dupuy}. Assuming a uniform
distribution of orbital eccentricities from 0 to 1, we multiply the
projected physical separation by 1.1$^{+0.9}_{-0.36}$ to estimate the
true semi-major axis.
Applying this factor to \cand's measured separation, we obtain an
estimated orbital semi-major axis of 83$^{+69}_{-33}$~AU. We derived a
mass of 0.4\msun\ for \prim\ using its age and absolute $H$ mag with
\citet{2000A&A...358..593S} evolutionary models. Using this mass with
Kepler's Third Law we obtain a period of 1195$^{+1490}_{-710}$ years
for component B. The observed motion of the companion (1.6~AU over 1
year) is plausible for a bound object.

\subsection {Near-IR Colors and Metallicity \label{sec:colors}}

To measure colors of \prim, we convert our relative flux ratio
measurements (\S~2.1.3 and \S~2.2) into apparent magnitudes for \cand\
using the 2MASS photometry. The $JHK$-band IRTF/SpeX data yielded more
accurate flux ratios than the NICI observations, because the IRTF
photometry did not incur the additional uncertainty of the NICI focal
plane mask transmission; thus, we use the IRTF-derived results for the
photometry of \cand. However, the NICI images provide far better 
astrometry than the IRTF data. Table~\ref{sysprops} summarizes the
resulting magnitudes for \cand.

\begin{figure}[ht] 
  \centerline{
    \includegraphics[width=9cm]{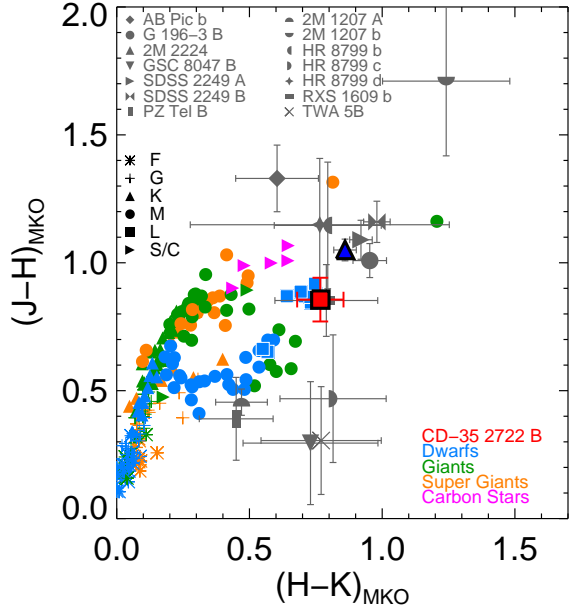}        
  }
  \caption{The infrared colors of \cand\ (center red square) plotted along with
    the colors of field objects and low mass companions. The $JHK$ colors of \cand\ are consistent with
    a late-type dwarf, but slightly bluer than that of 2MASS J22244381$-$0158521 (dark blue triangle), 
   the dusty L4.5 with very similar spectra. The colors also agree completely with those of 1RXS~J160929.1$-$210524~b, 
   which is plotted as a grey rectangle, almost covered by the symbol of \cand . 
   The colors of the field objects are derived from synthesized photometry
    of published spectra \citep[from][]{2009ApJS..185..289R}. The
    L~dwarf colors come from the compilation by \citet{2010ApJ...710.1627L}
    \citep[based on data from ][]{2002ApJ...564..466G, 2004AJ....127.3516G,
      2004AJ....127.3553K, 2006AJ....131.2722C}, with the plotted
    symbols representing the average color for each L~subclass. Also
    plotted for comparison are the NIR colors for known young late-M and L dwarfs
    (gray symbols; AB Pic b [\citealt{2005A&A...438L..29C}]
    , G~196-3 B [\citealt{1998Sci...282.1309R}]
    , 2MASS J22244381$-$0158521
    [\citealt{2004AJ....127.3553K}], GSC 8047-0232 B
    [\cite{2005A&A...430.1027C}], SDSS J053951.99$-$005902.0 AB
    [\citealt{2010ApJ...715..561A}], TWA 5 B
    [\citealt{1999ApJ...512L..69L}], 1RXS J160929.1$-$210524~b
    [\cite{2008ApJ...689L.153L}], 2MASS 1207334$-$393254 AB
    [\citealt{2007ApJ...657.1064M}], PZ Tel B
    [\citealt{2010ApJ.biller.pz}], HR 8799 bcd
    [\citealt{2008Sci...322.1348M}]). Where necessary, color
    transformations from the 2MASS to the MKO system were made using the
    \citet{2004PASP..116....9S} prescriptions for L dwarfs. }
  \label{jhk}
\end{figure}

Figure~\ref{jhk} compares the near-IR colors of \cand\ with other stellar and substellar
objects, including old (high-gravity) field ultracool dwarfs and young
(lower-gravity) objects found in the field and as companions. The $JHK$
colors of \cand\ agree well with known late-M and L~dwarfs. 

A metallicity measurement for the primary star is relevant to constraining the formation of \cand .
However, we see that \prim\ has [Fe/H] = 0.04\pp0.05~dex \citep{2009A&A...501..965V}, which is close to 
the mean [Fe/H] of the solar neighborhood, $-$0.05~dex \citep{2009ApJ...699..933J}.
As a consistency check, we make an indirect estimate of the metallicity of \prim\ using the photometric calibration of
\cite{2009ApJ...699..933J}.
Following their method, we determined the [Fe/H] of \prim\ from its
height above the main-sequence isochrone in a plot of $M_K$ versus
$V-K$. Since M1 dwarfs reach the main sequence by 100~Myr
\citep{2000A&A...358..593S}, the comparison of \prim\ with nearby
main-sequence stars is appropriate. 
We derived an [Fe/H] of
0.0\pp0.16~dex for \prim, consistent with the published value.
An alternate calibration of the metallicity is found in \citet{2010A&A...519A.105S}, 
who use a volume-limited and kinematically matched sample of F and G dwarfs from the Geneva-Copehnagen Survey to infer the 
mean metallicity of M dwarfs in the Solar Neighborhood and base their final calibration purely on high-resolution spectroscopy of FGK primaries with M dwarf companions.
Using their method, we derived an [Fe/H] of -0.12\pp0.16~dex for \prim , again 
consistent with a solar metallicity.

\subsection{Spectral Analysis \label{sec:spectra}}
\subsubsection{Spectral Typing \label{sec:spectyping}}

\begin{sidewaysfigure}
  \centerline{
    \hbox{
      \includegraphics[width=7cm]{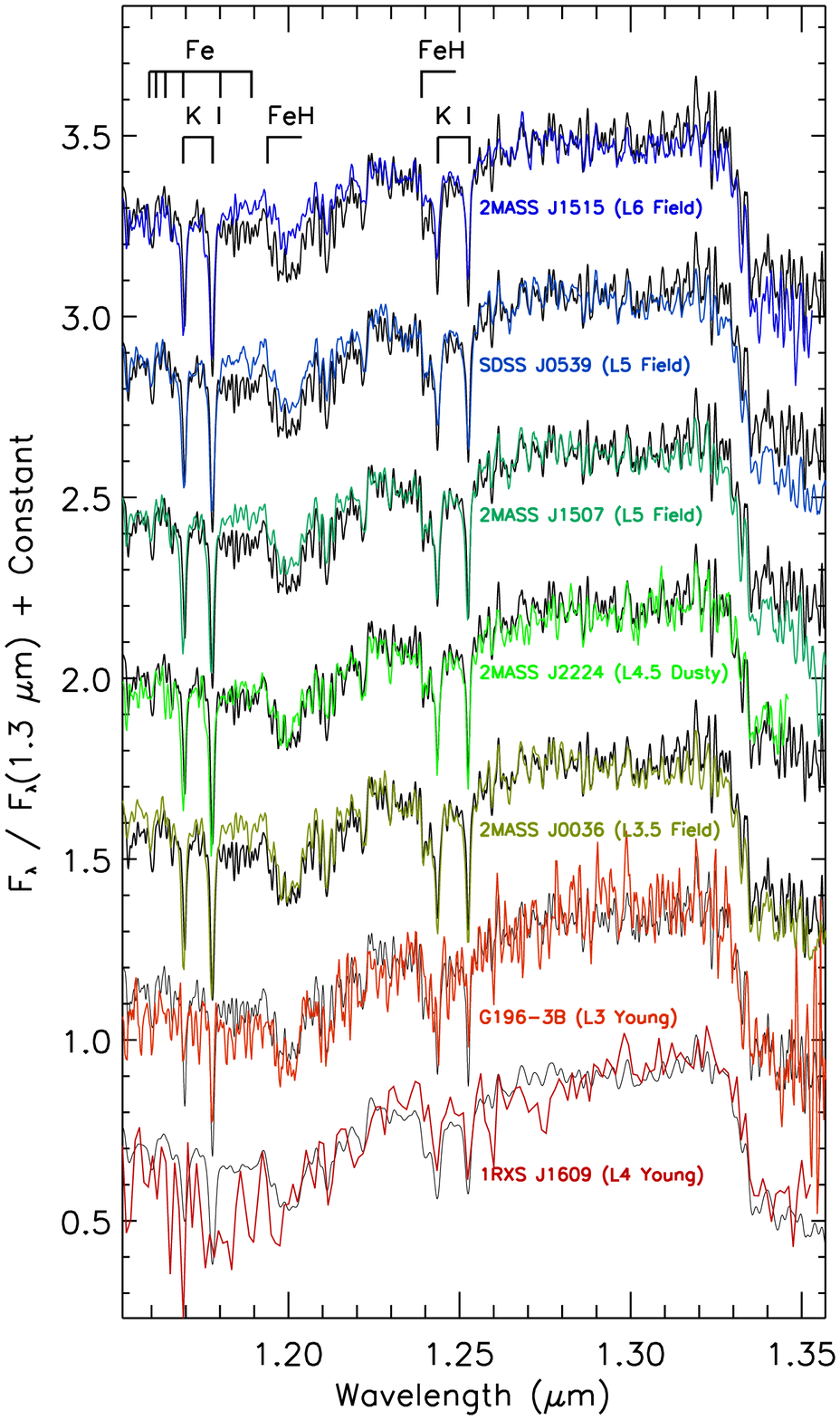}        
      \includegraphics[width=7cm]{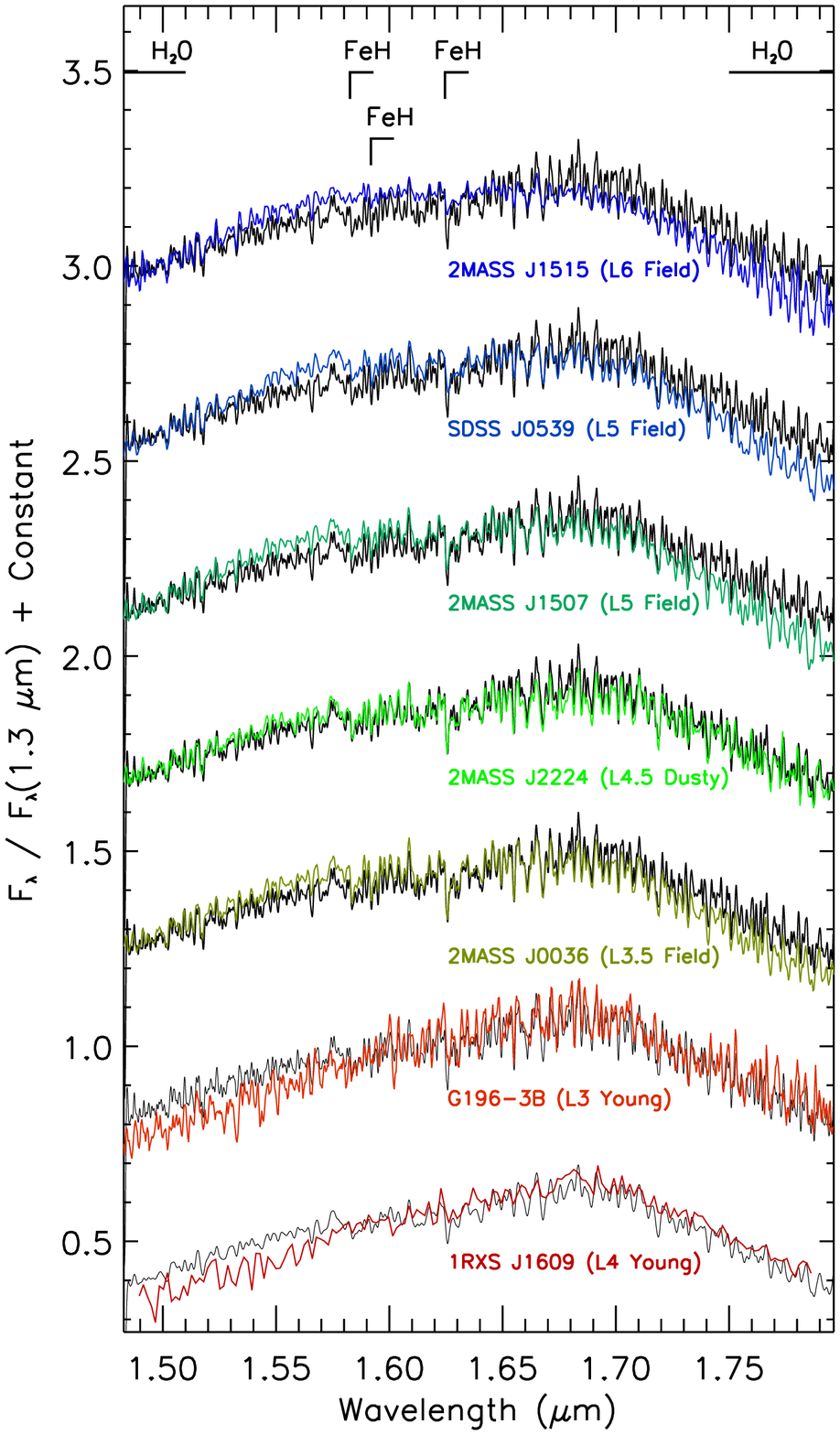}
      \includegraphics[width=7cm]{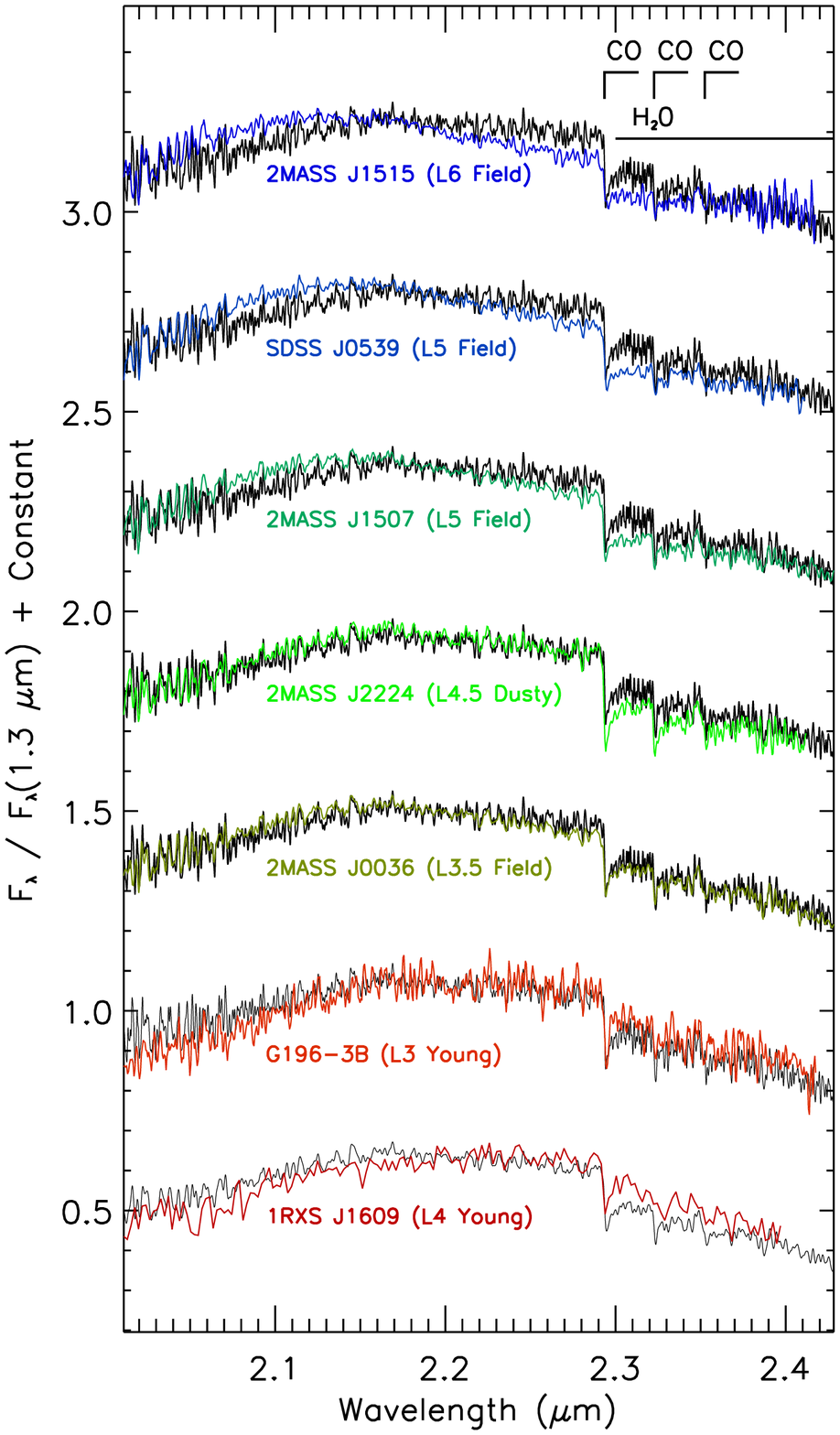}
    }
  }
  \caption{\scriptsize{Gemini NIFS spectra of \cand\ (black curves)
      compared to young (low gravity or dusty) and field L dwarfs. The
      $J$, $H$ and $K$-band spectra are shown in the left, middle and
      right panels, respectively. The NIFS spectrum is smoothed to match
      the resolution of the comparison spectra, which have been scaled
      to fit each band separately. The \cand\ spectrum is
      most similar to that of the dusty L4.5 2MASS J22244381--0158521
      and the low gravity L3 dwarf G196--3B, and begins to differ
      significantly from spectra which are more than one subclass away
      from L4. Thus we assign a spectral type of L4\pp1 to \cand. The
      top five comparison spectra (2MASS J15150083+4847416, SDSS
      J053951.99-005902.0, 2MASS J15074769-1627386, 2MASS
      J22244381-0158521, 2MASS J00361617+1821104) are from the IRTF
      spectral library \citep[$R=2000$;][]{2005ApJ...623.1115C}. The 1RXS J160929.1--210524~b spectra
      ($R=850$) are from \citet{2008ApJ...689L.153L}, while the G196--3~B
      spectra are from \citet{2010ApJ...715..561A}. Molecular features
      and line identifications are from \citet{2005ApJ...623.1115C}.} }
  \label{nifs_emp_jhk}
\end{sidewaysfigure}

To derive an empirical spectral type for \cand , we compared our
Gemini/NIFS spectra to field L dwarfs from the IRTF Spectral Library \citep{2005ApJ...623.1115C,2009ApJS..185..289R} and
young L dwarfs from recent studies (Figure~\ref{nifs_emp_jhk}).
When comparing to lower resolution spectra (IRTF prism R=150, IRTF SpeX
R=2000), we convolved our NIFS spectrum with a Gaussian of appropriate
width. Since the field dwarfs are old objects, we expect some mismatch with
\cand\ in surface gravity dependant spectral features. 

The best matched spectra we find is from the L4.5 dwarf 2MASS
J22244381$-$0158521, which is believed to have an unusually dusty
atmosphere \citep[see][]{2009ApJ...702..154S}.  
We note that almost every absorption
line feature is well-matched between this dusty field object and \cand.
However the NIR colors 
of \cand\ are a little bluer than those of 2MASS~J22244381$-$0158521 
($H-K$=0.77 vs 0.86 and  $J-H$=0.86 vs 1.05; see Figure~\ref{jhk}).
The young ($\approx$~5~Myr) L4 planet 1RXS~J160929.1$-$210524~b\citep{2008ApJ...689L.153L,2010ApJ...719..497L} 
exhibits a narrower $H$-band shape than the \cand\ spectra, but the $JHK$ absolute magnitudes and $J-H$ and $H-K$ colors 
of the two objects agree completely within errors. 

The field L3 and L5 dwarfs show clear differences in line strengths,
continuum shape and flux level in the $K$-band. The \cand\ spectrum also
shows signs of youth in that the shape of its $H$-band spectra is more
triangular than that of field L dwarfs, as seen in young L-type dwarfs
\citep{2006MNRAS.373L..60L, 2006ApJ...639.1120K, 2007ApJ...657..511A}
 However, the $H$-band shape is also less peaked
than the young L~dwarfs 1RXS~J160929.1-210524~b ($\sim$5~Myr; \citealt{2010ApJ...719..497L}) 
and G196-3~B ($\approx$~100~Myr; \citealt{1998Sci...282.1309R}, \citealt{2009ApJ...699..649S}).
Overall, \cand\ does not seem to have
especially low surface gravity, since the strength of its FeH lines and
K~I lines are comparable to those of field dwarfs.

Since the L3 and L5 spectra already differ significantly from that of
\cand, we assign \cand\ a spectral type of L4\pp1. 
We also derive consistent 
spectral types of L2.5, L3.5, L3.5 and L4 using spectral indices from
the literature: H$_2$O~1.53\micron\
\citep[1.27\pp0.05;][]{2007ApJ...657..511A}, H$_2$OA
\citep[0.56\pp0.03;][]{2003ApJ...596..561M}, H$_2$OC
  \citep[0.65\pp0.03;][]{2003ApJ...596..561M}, and CH$_4$~2.2\mic\
  \citep[0.946\pp0.005;][]{2002ApJ...564..466G}, respectively.
 
We also tried to find the best matched spectra to \cand\ from the $>$ 400 M, L, and T dwarfs in the
online archive of SpeX Prism data \citep[e.g.][]{2006AJ....131.1007B}, by scaling relative flux levels to minimize $\chi_\nu^2$.
To match the low-resolution SpeX spectra, our NIFS spectra are first
Gaussian smoothed. At this lower resolution, most features of the
spectra are no longer visible, and so fits rely mainly on the shape of
the continuum. Nevertheless, we observe a clear minimum in the
$\chi_\nu^2$ value for spectral types between L3 and L6. We verified the 
fits by eye to check that the $\chi_\nu^2$ minima 
were sensible. The closest matches to our smoothed spectra, in order of
increasing $\chi_\nu^2$ are 2MASS J20025073-0521524 (L6,
\citealt{2008ApJ...681..579B}), SDSS J232804.58-103845.7 (L3.5,
\citealt{2006AJ....131.2722C}), 2MASSW J0205034+125142 (L5, Reid et al.\
2006) and 2MASS J21512543 -2441000 (L3, \citealt{2008ApJ...681..579B}).

\begin{figure}[ht] 
  \centerline{
    \hbox {
      \includegraphics[width=20cm]{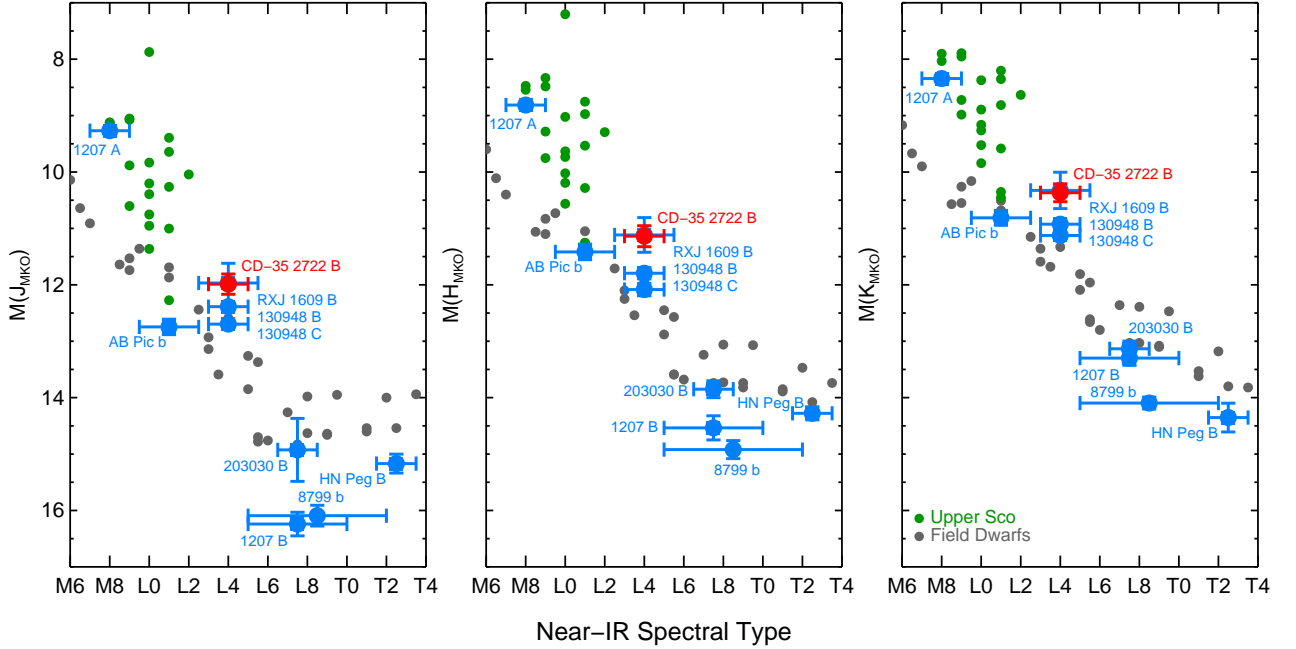}
    }        
}
\caption{ The absolute J, H and K MKO magnitudes and spectral type of \cand\ compared to cool field objects and other young MLT dwarfs
with parallax measurements. In the NIR, \cand\ is over-luminous compared to field objects, but has the same JHK luminosities as 
the 5 Myr old Upper Sco planet, 1RXS~J160929.1$-$210524~b. Earlier young dwarfs are generally also over-luminous, 
while later young dwarfs are under-luminous. The objects shown are the MLT dwarf compilation from \citet{2010ApJ...710.1627L} (gray symbols), 
Upper Sco L dwarfs from \cite{2008MNRAS.383.1385L} (green symbols), 
2MASS~1207334$-$393254~AB \citep{2007ApJ...657.1064M}, 
AB Pic~b \citep{2005A&A...438L..29C},
1RXS~J160929.1$-$210524~b \citep{2008ApJ...689L.153L}, 
HD~130948~BC \citep{2009ApJ...692..729D},
HD 203030~B \citep{2006ApJ...651.1166M},
HR~8799~b \citep{2008Sci...322.1348M,2010ApJ...723..850B} and
HN Peg b \citep{2007ApJ...654..570L}, all plotted using blue symbols.}
\label{lumcomps}
\end{figure}

With a spectral type estimate for \cand\ in hand, we compare
its NIR luminosities with those of field MLT dwarfs, and other young MLT dwarfs with measured 
parallaxes (see Figure~\ref{lumcomps}). \cand\ is over-luminous by \app\ 1~mag 
compared to L4 field dwarfs in the JHK-bands. As noted earlier, its NIR luminosities are indistinguishable 
from those of  the 5~Myr old planet, 1RXS~J160929.1$-$210524~b. The young L4 dwarfs 130948~B and C (age \app\ 790 Myr, \citealt{2009ApJ...692..729D})  
are also over-luminous by \app\ 0.5 mags. 
Young late-L and early-T companions are
clearly under-luminous compared to ﬁeld objects, as previously noted by
\citet{2006ApJ...651.1166M} and \citet{2010ApJ...723..850B}, while the young
late-M to mid-L companions are overluminous.
The young L1 dwarf, AB Pic b (age \app\ 30 Myr, \citealt{2008hsf2.book..757T}) 
seems under-luminous and thus unlike other young M7--L5 dwarfs.   

\subsubsection{Model Atmosphere Fitting \label{sec:maf}}
\begin{figure}
  \centerline{
    \includegraphics[height=9cm]{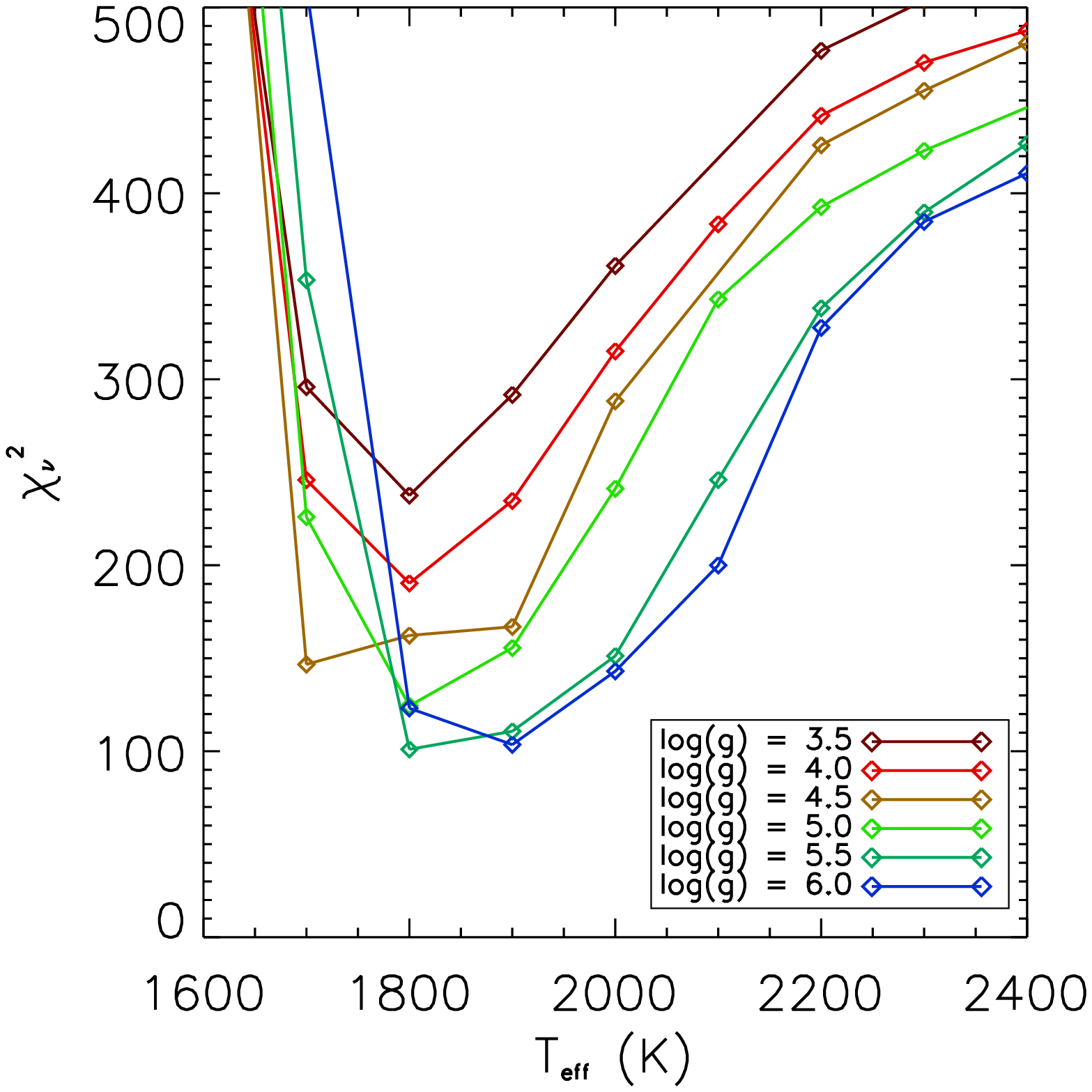}        
      \includegraphics[height=9cm]{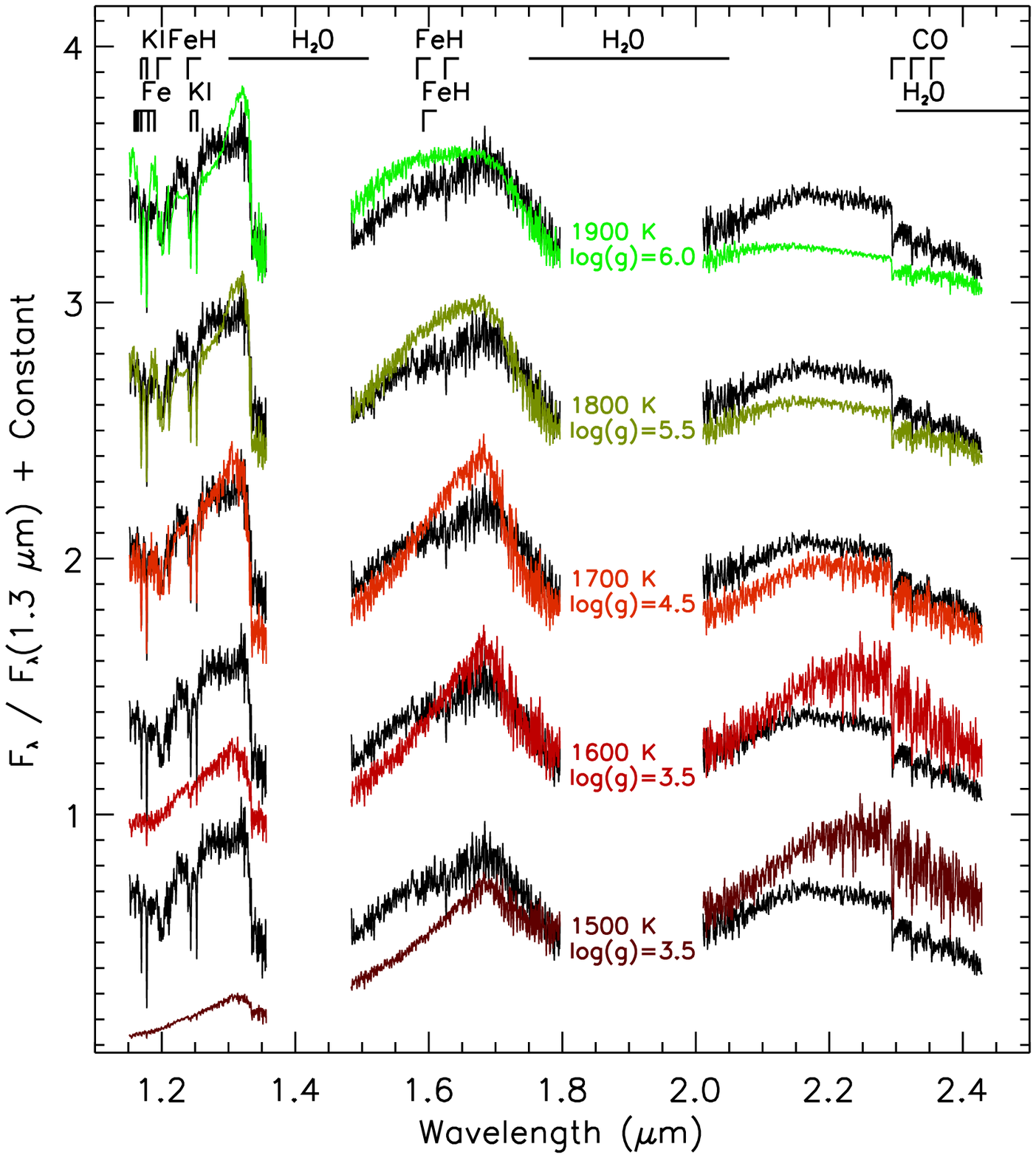}
  }
  \caption{Left: The $\chi_\nu^2$ from fitting Ames-Dusty model
    atmospheres \citep{2001ApJ...556..357A} to the NIFS spectra   
of \cand\ as a function of model temperature and gravity. The best fit models
are the T$_{eff}$ = 1700--1900~K models with $\logg = 4.5-6.0$~dex. However,
none of the models match the continuum shape very well.
The reduced chi-squared values quickly deteriorate for T$_{eff} \leq$ 1600~K
and T$_{eff} \geq$ 2000~K models. 
Right: The best fit Ames-Dusty model spectra for different effective
temperatures. }
  \label{ames_chi2}
\end{figure}

To constrain the atmospheric properties of \cand , we compare our NIFS
spectrum to the Ames-Dusty synthetic spectra of
\citet{2001ApJ...556..357A}. For each model spectrum in the Dusty grid,
the flux of the model is adjusted to minimize the $\chi_\nu^2$ value of
the fit to the observed spectrum. The overall minimum $\chi_\nu^2$ value
across all template spectra in the grid gives the best-fit model
(Figure~\ref{ames_chi2}).
We find the smallest $\chi_\nu^2$ values for $T_{eff}$ = 1700--1900~K with
log(g) = 4.5--6.0. Since, Ames-Dusty models are not able to reproduce very well the 
spectra of cooler objects near the L-T transition, the lower temperature bound should 
not be taken as a very strong constraint. Nevertheless, using models from \citet{2008ApJ...689.1327S}, \citet{2008ApJ...678.1372C} are
able to fit the spectra of L4 and L5 dwarfs with their \Teff=1700~K, log(g)=4.5 models and L6 dwarfs with their \Teff=1400~K, log(g)=4.5 
atmospheric models, achieving good-fits past spectral type L8. This suggests that 
our lower bound of \Teff=1700~K is reasonable. Our temperature estimate is also consistent with the temperature
we derive from the Lyon/Dusty evolutionary models in \S~\ref{sec:masses}.

 Since $\chi_\nu^2$ values depend more on the quality of matches between the shapes
of the continuum rather than matches between specific spectral features, we also judge the best-fit models
by eye. 
Examining the fits, we note that while no single model is 
a perfect match to the \cand\ spectra, the 1700K, log(g) = 4.5 model reproduces the overall continuum shape from $J$ to $K$ rather well (Figure~\ref{ames_chi2}).
At the same time, the shape and the depth of the two sets of K~I doublets at 1.17 and 1.25
$\mu$m and the FeH lines in the $J$-band are clearly better reproduced in the 1700~K model than in other models. 
Moreover, given the young age of the system, we expect it to have a low
surface gravity. The lower gravity of the 1700~K, log(g) = 4.5 model is also consistent with the value estimated from the
Lyon/Dusty evolutionary model using our bolometric luminosity and age
estimates (\S~\ref{sec:masses}). Since the quality of the J-band fits are clearly much worse for the 1700~K, log(g)=4.0 and log(g)=5.0 models,
we estimate a log(g)=4.5$\pm$0.5 for \cand .

We also computed $\chi_\nu^2$ from the fits to L dwarf models with clouds presented in \citet{2006ApJ...640.1063B}.
The shapes of the continua and the relative fluxes in the $JHK$-bands are completely consistent between these models and the \cand\ spectra. The 
strengths of the K~I doublets in the $J$-band are also consistent.
The $\chi_\nu^2$ results show a clear minima in the \Teff--log(g) space, and we estimate T$_{eff}=$1700$\pm$100~K, and log(g)=4.5--5.0 from the Burrows models. 
The $H$-band peaks in the models are, however, clearly rounder with more flux at the blue end compared to the \cand\ spectra due to the absence of 
FeH features in the models.
The models also show stronger water absorption at the red end of the $H$-band.
Moreover, the models are deficient in Fe and FeH features in the $J$-Band, and their CO features in the $K$-band are too strong.
The cloud-free models from \citet{2006ApJ...640.1063B} have very different continua shape from the \cand\ spectra, and a reasonable match was not found.

\subsubsection{Radial Velocities \label{sec:rv}}
Due to the dither pattern in the first set of $J$-band observations,
there was enough light in the NIFS data to extract a spectrum of \prim .
We used this spectrum to test whether \prim\ and B share the same radial
velocity. The spectral types of the two objects are sufficiently
different that cross-correlation techniques are not easily applied.
Instead, we fit Gaussians to four strong K~I lines in the $J$ band,
taking the average of the centroids as our measure of radial velocity.
We find that the velocities of A and B agree to within the errors (\pp10
km/s) further confirming their physical association. The system velocity
(25\pp10~km/s) also agrees with the published value \citep[31.4\pp0.4
km/s,][]{2006A&A...460..695T} and the predicted value in \S~3.1 within our errors.

\subsection{Mass Estimates \label{sec:masses}}

We first determined the mass of \cand\
using the Lyon/Dusty evolutionary models \citep{2000ApJ...542L.119C},
based on the observed absolute magnitudes and estimated age. To compute
the mass and its associated uncertainty, we used a Monte Carlo
technique, drawing the input values for $M_J$ and age from random
distributions. Values for $M_J$ were chosen from a normal distribution
corresponding to the photometry in Table~\ref{sysprops}. Age values were
chosen from a uniform distribution between 50 and 150~Myrs, as estimated
for the AB~Doradus association \citep{2004ApJ...613L..65Z,
  2005ApJ...628L..69L, 2007ApJ...665..736C}. Using the Lyon/Dusty
evolutionary models in the MKO photometric system (I. Baraffe, private 
communication), we computed the mass of \cand\ from 10,000 random
trials, resulting in an adopted mass of 31\pp7~\mjup\ from the median
and standard deviation of the trials. Using the same approach with our
$M_H$ and $M_K$measurements, we estimate masses of 30\pp7 and 34\pp9~\mjup ,respectively.
Thus, \cand\ is probably a brown dwarf, if we follow the notional dividing line between planets and brown dwarfs set by 
the deuterium-burning limit of 13.6 \mjup\ for solar metallicity.

To assess the uncertainty contributed by each input parameter (age,
distance, and $J$ mag) in our mass calculation, we also computed the
mass of \cand\ by assuming two of the inputs were error-free. The mass
uncertainty arising from the assumed age dominates over that from the distance, 
which in turn dominates over that from the $J$-band photometry (Table~\ref{errmass}).
Of course, this does not account for the systematic errors in the
evolutionary models themselves, which are just now being rigorously
tested \citep[e.g.][]{2007ApJ...664.1154S,2009arXiv0912.0738D}.

\begin{figure}[ht] 
  \centerline{
    \hbox {
      \includegraphics[width=6cm]{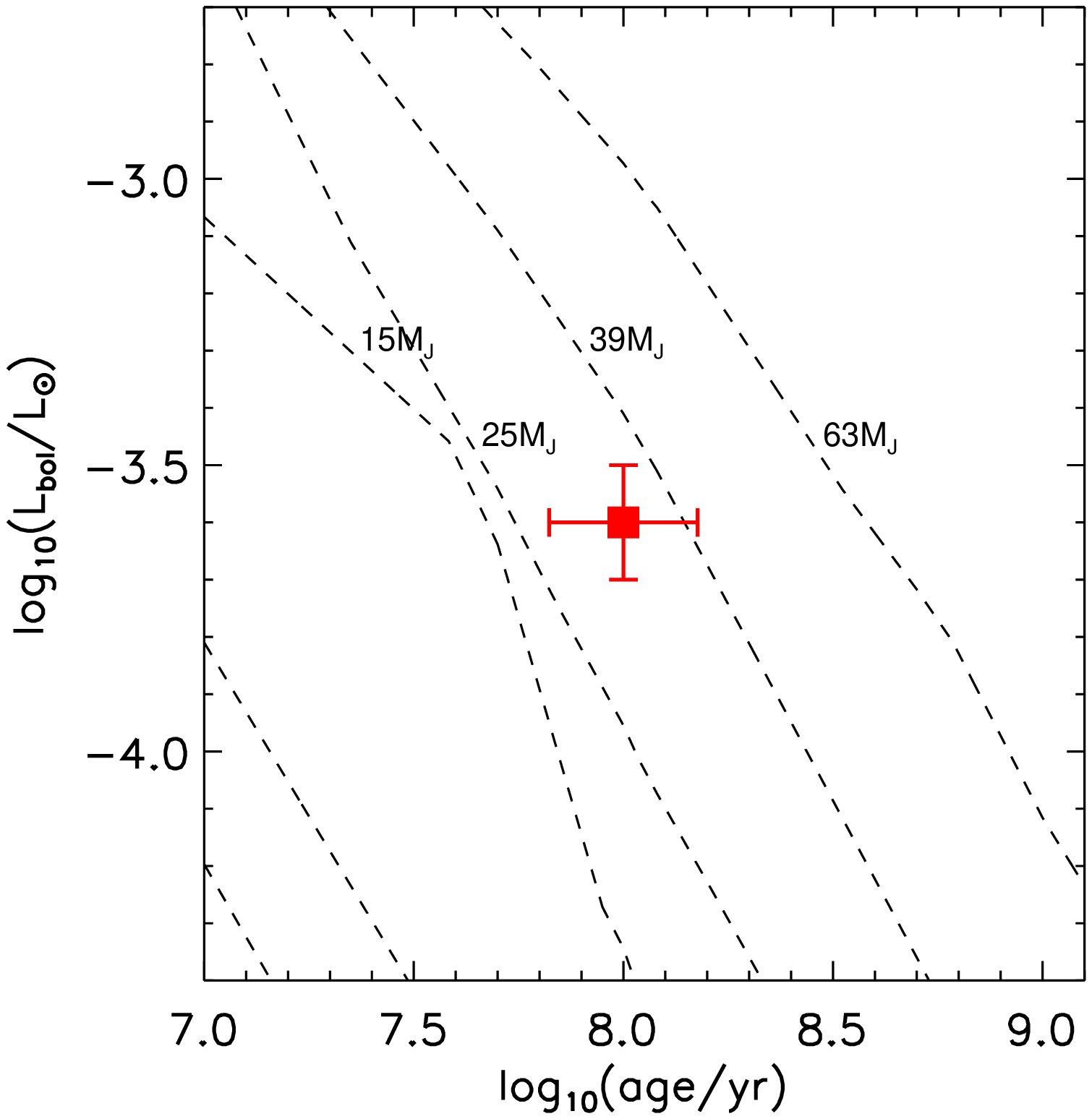}
      \includegraphics[width=6cm]{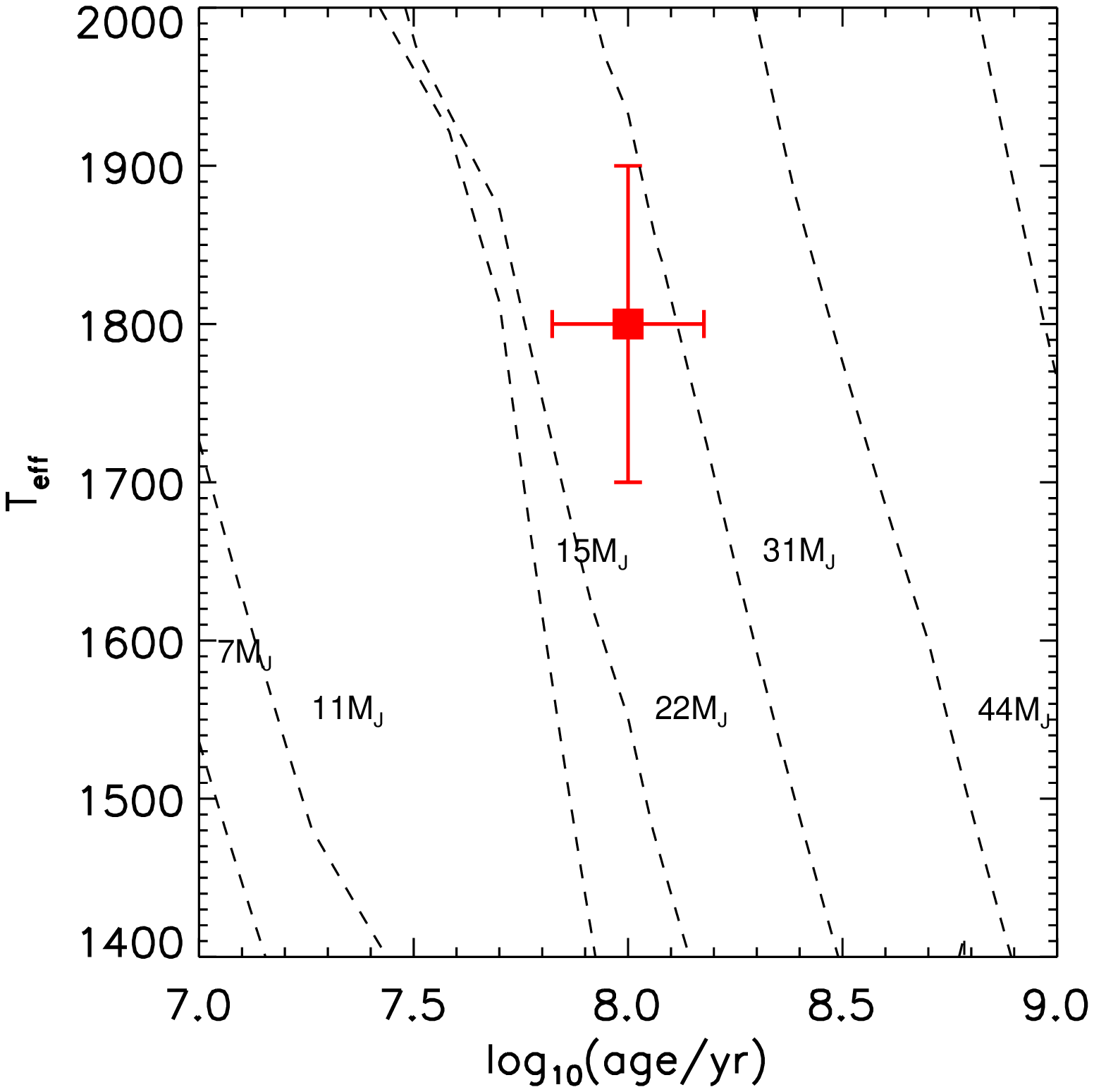}
      \includegraphics[width=6cm]{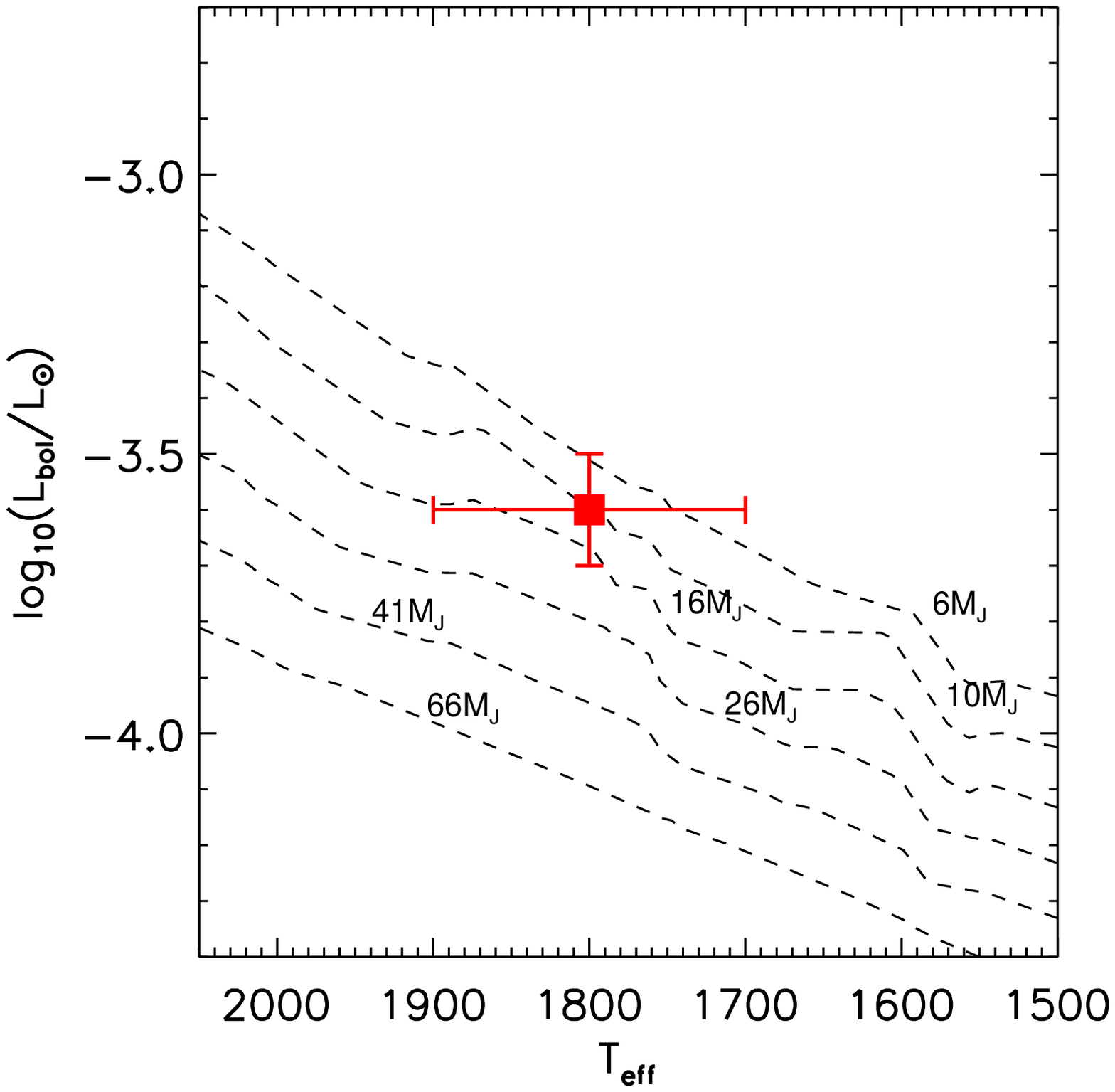}
    }        
}
\caption{ Comparison of the properties of \cand\ to those predicted by
  the Dusty model mass tracks. The mass estimate derived from the
  bolometric luminosity and age and from the \teff\ (derived from model
  atmosphere fits to the near-IR spectra) and age are consistent with
  each other. }
\label{massests}
\end{figure}

A somewhat more robust way to derive the mass of \cand\ is to use its
bolometric luminosity and age as inputs to the Lyon/Dusty evolutionary
models, since bolometric luminosities are less subsceptible to
uncertainties in the model atmospheres than single bandpass fluxes
\citep{2000ApJ...542L.119C}.
Using the $J$-band bolometric corrections from Liu et al.\ (2010) for
spectral types of L3--L5 (\S~\ref{sec:spectra}) and the input
distribution of $M_J$ described above, we calculated the corresponding
distribution of bolometric luminosities for \cand. Then, using the
luminosity and age distributions with the evolutionary models, we
obtained a mass of 31\pp8\mjup, again using the median and standard
deviation of the trials (Figure~\ref{massests}). The $L_{bol}$-derived masses 
from the $H$ and $K$-bands were also 31\pp8\mjup , which is expected since 
most of the mass uncertainty arises from the uncertainty in age. 
We adopt this $L_{bol}$-derived mass as our best estimate in
Table~\ref{sysprops}, though the mass estimates from individual filters
and from $L_{bol}$ are completely consistent. Note that the same set of input data
and evolutionary models give an effective temperature of
1980\pp100~K 
and a surface gravity with $\log(g)=
4.54\pm0.13$~dex, 
in reasonable agreement with the results from model atmosphere fits to the near-IR
spectra (\S~\ref{sec:spectra}). 

Using the same methods as for the Lyon/Dusty models, we compare the age and luminosity of \cand\ to 
the models of \citet{1997ApJ...491..856B,2001RvMP...73..719B} to estimate the mass, 
\Teff , and $log(g)$ to be 36~\pp6\mjup , 2150\pp90~K and 4.83\pp0.13 dex, respectively.

An alternative means using evolutionary models to estimate the mass
comes from using the age and effective temperature, with the latter estimated from
model atmosphere fitting of the NIR spectra (\S~\ref{sec:spectra}).
We adopt a similar Monte Carlo approach, assuming a
  uniform distribution in \Teff\ from 1700--1900~K, and derive a mass of
  $23\pm5$~\mjup. This is somewhat lower than the mass derived from
\{\Lbol, age\}. This discrepancy occurs because the model
atmosphere-derived \Teff\ is $\approx$~180~K cooler than the evolutionary
model-derived \Teff\ above, leading to a lower mass.\footnote{A somewhat
  similar effect has been identified in the young ($\approx$~790~Myr)
  L4+L4~dwarf binary HD~130948BC, where the \Teff\ derived from
  evolutionary models using the age and luminosity are
  $\approx$~150--300~K hotter than expected from atmospheric model
  fitting for objects of similar spectral type. (A direct model fit to
  the HD~130948BC spectrum is not available.) See \S~5.8 of \citealt{2009ApJ...692..729D}.}

A third means to estimate the mass from the evolutionary models comes
from the H-R diagram, namely using the bolometric luminosity and the estimated temperature. 
This is the least useful method, as small uncertainties in \Teff\ translate into large mass uncertainties,
given the tightness of the model tracks on the H-R diagram. Assuming a
normal distribution in luminosity and a uniform distribution in \Teff, 
we derive $11 \pm 7$~\mjup. Figure~\ref{massests}
illustrates the observed properties of \cand\ relative to the model
predictions for different pairs of observable properties.

\begin{figure}[ht] 
  \centerline{
    \hbox {
      \includegraphics[height=9cm]{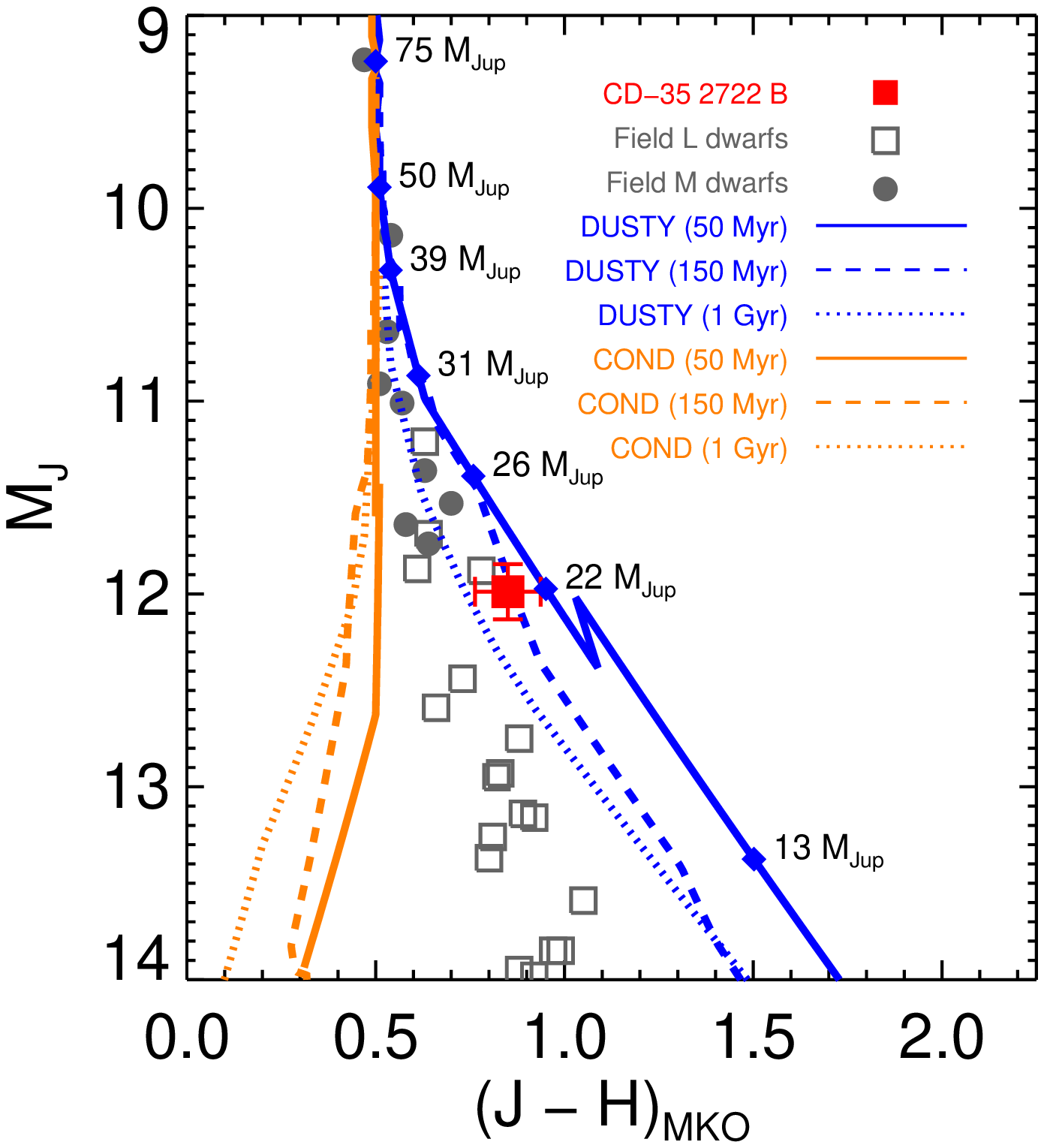}
      \includegraphics[height=9cm]{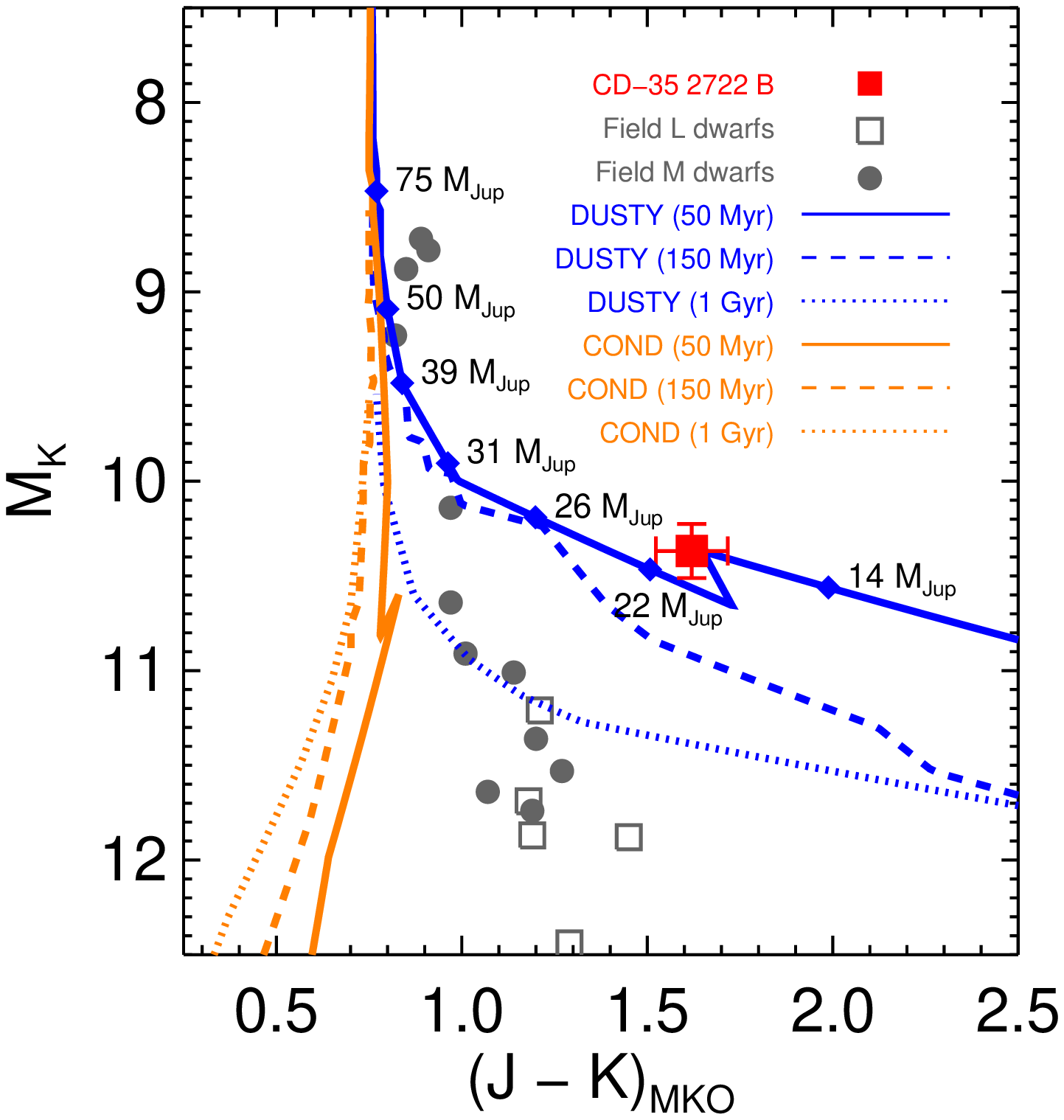}
    }        
}
\caption{Comparison of the COND and Dusty model isochrones at 50 Myr, 150
  Myr and 1 Gyr with the absolute magnitudes and NIR colors of \cand\ shown by
  the red square. For comparison, we also plot the NIR magnitudes and colors of field M and L~dwarfs, 
  from a compilation by Sandy Leggett \citep{2002ApJ...564..466G, 2004AJ....127.3516G,
      2004AJ....127.3553K, 2006AJ....131.2722C}. As is well known, the COND and Dusty models do not predict the 
  field brown dwarf sequence well at low temperatures.}
  \label{cmd}
\end{figure}

The color-magnitude diagrams in Figure~\ref{cmd} compare the 50 and
150~Myr isochrones from the Lyon/Dusty and Lyon/COND
\citep{2003IAUS..211...41B} evolutionary models with the color and
absolute magnitudes of \cand. In both plots, the $M_J$ vs $J-H$ and the
$M_K$ vs $J-K$ diagram, the companion's position is within 1$\sigma$ of
that of 22, 31 and 39~\mjup\ Dusty model-simulated brown dwarfs at ages 50, 100 and 150~Myrs, respectively.

\section {Conclusions}

As part of the Gemini NICI Planet-Finding Campaign, we have discovered a
substellar companion to the young M1V star \prim, a member of the AB
Doradus association. 
High quality astrometry over one year confirms \cand\ as a bound
companion, with a projected separation of 3.17\arcsec\ (67$\pm$4~AU) in
January 2009. We estimate a companion mass of 31\pp8~\mjup\ by using its
bolometric magnitude and age and the Lyon/Dusty evolutionary models.
$JHK$ photometry show that the system has L~dwarf-like colors. 

The 1.1--2.4~\micron\ spectra of \cand\ reveal a low temperature and low surface gravity,
as expected for the young age and low mass of the companion. The
$H$-band continuum shows the triangular shape characteristic of young
ultracool dwarfs, though intermediate in shape between the youngest
known objects and old (field) objects. Through direct comparison with
known objects, we derive a spectral type of L4\pp1, making \cand\ one of
the coolest young substellar companions identified to date. We fit the
spectra using the Ames-Dusty model atmospheres and derive
$T_{eff}=$1700-1900~K and $\log(g) = $4.5$\pm$0.5~dex. These estimates are consistent with the 
Lyon/Dusty evolutionary model-derived temperature of 1980\pp100~K  and $\log(g) = $4.54$\pm0.14$~dex.
As a member of a moving group with a good age estimate, \cand\ serves as a 
valuable benchmark for calibrating age-dependent spectral effects in
L~dwarfs. Comparison with other young substellar
companions shows that when using near-IR spectral types, the late-M to
mid-L companions appear to be over-luminous compared to the field
objects, while the late-L and early-T companions appear to be
under-luminous.

The mass and separation of \cand\ are comparable to other known young
star+substellar companion systems. TWA~5b (20 \mjup, 98 AU; Lowrance
1999), GQ Lup B (17 \mjup, 100 AU; Neuhauser et al.\ 2005) and Gl~229~B
(35~\mjup, 45~AU; Nakajima et al.\ 1995) all have masses and separations
within 50\% of that of \cand, and all have primaries of spectral type K7
or later. This similarity may be a hint that they share a common
formation mechanism. In-situ formation by core accretion is not viable
for such massive companions at such large separations, especially around
an M star. Alternate theories do not provide easy pathways to the
formation of substellar companions like \cand; for instance, formation
by fragmentation of a prestellar core (Bate 2003) is difficult for the
mass ratio and separation of
\syst~AB.
Formation via gravitational instability (Boss 1997, Rafikov 2005, Boss
2006, Stamatellos \& Whitworth 2008) require large massive disks
($>$0.2-0.35~\msun) and thus also do not provide an obvious solution,
given the low mass of the primary star. However, Jeans mass
fragmentation of the interstellar molecular cloud may be able to explain
stellar-substellar systems, \eg, as discussed by Zuckerman and Song
(2009). Moreover, \citet{2001ApJ...551L.167B} found that dense molecular
cloud cores with magnetic fields can collapse and fragment into multiple
protostar systems with initial masses as low a Jupiter mass, i.e., even
smaller than the companion to \prim. This suggests that the CD-35 2722
system might well have formed by the classic binary star formation
mechanism of fragmentation of a collapsing dense molecular cloud core.

\acknowledgements

This work was supported in part by NSF grants AST-0713881 and
AST-0709484. 
Our research has employed the 2MASS data products; NASA's Astrophysical
Data System; the SIMBAD database operated at CDS, Strasbourg, France;
the M, L, and T~dwarf compendium housed at DwarfArchives.org and
maintained by Chris Gelino, Davy Kirkpatrick, and Adam Burgasser; and
the SpeX Prism Spectral Libraries maintained by Adam Burgasser at
http://www.browndwarfs.org/spexprism. We would also like to thank Guillem Anglada and
Alycia Weinberger for their parallax measurements of \prim .

{\it Facilities:} Gemini-South (NICI), IRTF (SpeX).

\vfill
\eject

\bibliographystyle{apj}
\bibliography{zrefs}

\vfill
\eject

\clearpage
\begin{deluxetable}{lcc}
\tablecaption{Properties of the CD-35 2722 AB system. \label{sysprops}}
\tablewidth{0pt}
\tablehead{
 \colhead{Property} & 
 \colhead{\prim} &
 \colhead{\cand} }

\startdata

Age (Myr)\tablenotemark{a}                    & \multicolumn{2}{c}{100\pp50} \\
Distance (pc)\tablenotemark{b}                     &\multicolumn{2}{c}{21.3\pp1.4}\\
Angular separation (\arcsec)\tablenotemark{c}  & \multicolumn{2}{c}{3.137\pp 0.005} \\
Physical separation (AU)\tablenotemark{c}      & \multicolumn{2}{c}{67\pp 4} \\
Position angle (\dg)\tablenotemark{c}          & \multicolumn{2}{c}{243.10\pp 0.25}\\
$\Delta J$, IRTF (mag)                         &\multicolumn{2}{c}{5.77\pp0.06}\\ 
$\Delta H$, IRTF (mag)                         &\multicolumn{2}{c}{5.50\pp0.06}\\
$\Delta K$, IRTF (mag)                         &\multicolumn{2}{c}{4.98\pp0.05}\\
$\Delta J$, NICI (mag)                         &\multicolumn{2}{c}{5.66\pp0.12}\\ 
$\Delta H$, NICI (mag)                         &\multicolumn{2}{c}{5.48\pp0.15}\\
$\Delta K_S$, NICI (mag)                         &\multicolumn{2}{c}{4.96\pp0.07}\\
$\Delta K$, NICI (mag)\tablenotemark{d} &\multicolumn{2}{c}{4.94\pp0.07}\\
$J_{MKO}$ (mag)\tablenotemark{e}                &7.86\pp0.03&13.63\pp0.11\\
$H_{MKO}$ (mag)\tablenotemark{e}                &7.28\pp0.03&12.78\pp0.12\\
$K_{MKO}$ (mag)\tablenotemark{e}                &7.03\pp0.05&12.01\pp0.07\\
$M_J$ (mag)                                    &6.0\pp0.14&11.99\pp0.18\\
$M_H$ (mag)                                    &5.4\pp0.14&11.14\pp0.19\\            
$M_K$ (mag)                                    &5.1\pp0.14&10.37\pp0.16\\
$(J-H)_{MKO}$ (mag)                            &0.58\pp0.03&0.86\pp0.13\\
$(H-K)_{MKO}$ (mag)                            &0.25\pp0.02&0.77\pp0.2\\
Spectral Type                                  & M1Ve\tablenotemark{f}& L4\pp 1\\                  
M$_{bol}$ (mag)                                 &  8.11\pp0.25  &13.7\pp0.3 \tablenotemark{g}\\
Mass                                           & 0.4\pp0.05 \msun \tablenotemark{h}& 31\pp8 \mjup \tablenotemark{i}\\
T$_{eff}$ (K)                                   & 3680\pp100 \tablenotemark{j}&1700--1900 \tablenotemark{k}\\
\logg\ (dex)                                    & \nodata & 4.5$\pm$0.5 \tablenotemark{l}\\ 
$[$Fe/H$]$                                           & 0.04\pp0.05~dex \tablenotemark{m} & \nodata\\

\enddata

\tablenotetext{a}{Age estimate for the AB~Doradus association
  \citep{2004ApJ...613L..65Z,2005ApJ...628L..69L,2007ApJ...665..736C}.}

\tablenotetext{b}{Distance from parallax measurement using CAPScam. See \S~3.1 for details.}

\tablenotetext{c}{Epoch January 10, 2010 UT.}

\tablenotetext{d}{Transformed from NICI $\Delta K_S$ magnitude, using the relationship in \citet{2004PASP..116....9S}
and asumming a spectral type of L4.}

\tablenotetext{e}{MKO magnitudes derived from 2MASS magnitudes using
  color transformations in \citet{2006MNRAS.373..781L}. All photometry and colors are given in the MKO
  system. Apparent and absolute magnitudes, colors, and mass estimates
  for \cand\ are all derived from IRTF photometry. }

\tablenotetext{f}{Spectral Type from \citet{2006A&A...460..695T}.}

\tablenotetext{g}{Bolometric magnitude of primary derived using bolometric magnitude equations for M dwarfs in \citet{2008MNRAS.389..585C}. Bolometric magnitude for companion derived from $M_J$, based on bolometric corrections in Liu et al.\ (2010).}

\tablenotetext{h}{Mass estimate based on \citet{2000A&A...358..593S}
  using $M_H$ and age.} 

\tablenotetext{i}{Mass estimate based on Lyon/Dusty model using
  $M_{bol}$ and age.} 

\tablenotetext{j}{T$_{eff}$ for an M1V from \citet{2007ApJ...667..527G}}

\tablenotetext{k}{T$_{eff}$ from Ames-Dusty atmospheric model fitting result.}

\tablenotetext{l}{\ Log(g) from fitting Ames-Dusty model atmospheres to
  the Gemini/NIFS near-IR spectra.}

\tablenotetext{m} {[Fe/H] measured by \citet{2009A&A...501..965V}. }

\end{deluxetable}

\begin{deluxetable}{lccc}
\tablecaption{Astrometry of objects near \prim . }
\tablewidth{0pt}
\tablehead{
\colhead{Object} &\colhead{UT Date (Y/M/D)}&\colhead{Separation
  (\arcsec)} &\colhead{Position Angle (\dg)}}
\startdata
\cand &2009/01/06&3.172$\pm$0.005&244.13$\pm$0.25\\
  &2010/01/10&3.137$\pm$0.005&243.10$\pm$0.25\\ 
background object &2009/01/06&5.312$\pm$0.011&275.48$\pm$0.25\\
 &2010/01/10&5.316$\pm$0.011&276.32$\pm$0.25
\enddata
\label{astrom}
\end{deluxetable}

\begin{deluxetable}{ccccl}
\tablecaption{Spectroscopic observations obtained with Gemini-North/NIFS.}
\tablewidth{0pt}
\tablehead{ \colhead{UT Date (Y/M/D)} &\colhead{Band} & \colhead{No. of Exp.} & \colhead{Exp. Time(s)} & Comments }
\startdata
2010/03/07  & $J$ & 12 & 300 & \\
2010/03/18  & $H$ & 9 & 300 & Partial cloud\\
& $K$ & 4 & 120 \\
2010/03/19  & $J$ & 9 & 300 & Poor conditions, partial cloud
\enddata
\label{spectroscopy_observations}
\end{deluxetable}

\begin{deluxetable}{lcc}
\tablecaption{Comparison of the sources of uncertainty in \cand 's estimated mass.}
\tablewidth{0pt}
\tablehead{
\colhead{Parameter} & \colhead{Measurement} & \colhead{Derived Mass Error}
}
\startdata
Age & 100\pp50 Myr & 7 \mjup \\
Distance& 21.3\pp1.4~pc  & 0.6 \mjup \\
Apparent J mag & 13.63\pp0.11 mags & 0.3 \mjup \\\hline
All& \nodata & 8 \mjup 
\enddata
\tablecomments{For each parameter, we set the uncertainty in all the other 
parameters to zero and calculated the error in \cand 's mass using the Lyon/Dusty evolutionary models.}
\label{errmass}
\end{deluxetable}

\end{document}